\documentclass{svmult}

\usepackage{makeidx}         
\usepackage{graphicx}        
\usepackage{multicol}        
\usepackage[bottom]{footmisc}

\usepackage{amssymb}
\usepackage{amsfonts}
\usepackage{amscd}
\usepackage{amsmath}

\makeindex             

\begin{document}

\title*{Notes on the Self-Reducibility of the Weil Representation and
Higher-Dimensional Quantum Chaos}
\titlerunning{Self-Reducibility and Quantum Chaos}
\author{Shamgar Gurevich$^*$ and Ronny Hadani$\dagger$}
\institute{$^*$ Department of Mathematics, University of California, Berkeley, CA
94720, USA
\texttt{shamgar@math.berkeley.edu} \\
$\dagger$ Department of Mathematics, University of Chicago, IL, 60637, USA
\texttt{hadani@math.uchicago.edu}}
\authorrunning{S. Gurevich and R. Hadani}
%
\maketitle

\begin{abstract}
In these notes we discuss the \textit{self-reducibility property} of the
Weil representation. We explain how to use this property to obtain sharp
estimates of certain higher-dimensional exponential sums which originate
from the theory of quantum chaos. As a result, we obtain the Hecke quantum
unique ergodicity theorem for a generic linear symplectomorphism $A$ of the
torus $\mathbb{T=%
\mathbb{R}
}^{2N}/%
\mathbb{Z}
^{2N}.$
\end{abstract}


\section{Introduction}

\subsection{The Weil representation}

In his celebrated 1964 Acta paper \cite{W} Weil constructed a certain
(projective) unitary representation of a symplectic group over a local field $k
$ (for example $k$ could be $%
\mathbb{R}
,$ $%
\mathbb{C}
$, or a $p$-adic field). This representation has many fascinating properties
which have gradually been brought to light over the last few decades. It now
appears that this representation is a central object, bridging various
topics in mathematics and physics, including number theory, the theory of theta
functions and automorphic forms, invariant theory, harmonic analysis,
and quantum mechanics. Although it holds such a fundamental status, it is satisfying
to observe that the Weil representation already appears in the study of functions on
linear spaces. Given a $k$-linear space $L$, there exists an associated (polarized)
symplectic vector space $V=L\times L^{\ast }$. The Weil representation of the group $Sp=Sp(V,\omega )$
can be realized on the Hilbert space $\mathcal{H=}$ $L^{2}(L,%
\mathbb{C}
).$
Interestingly, some elements of the group $Sp$ act by certain kinds of
generalized Fourier transforms. In particular, there exists a specific
element $\mathrm{w}\in Sp$ (called the Weyl element) whose action is
given, up to a normalization, by the standard Fourier transform. From this
perspective, the classical theory of harmonic analysis seems to be devoted
to the study of a particular operator in the Weil representation.

In these notes we will be concerned only with the case of the Weil
representations of symplectic groups over finite fields. The main technical
part is devoted to the study of a specific property of the Weil
representation---the \textit{self-reducibility property}. Briefly, this is a
property concerning a relationship between the Weil representations of
symplectic groups of different dimensions. In parts of these notes we devoted some
effort to developing a general theory. In particular, the results concerning
the self-reducibility property apply also to the Weil representation over
local fields.

We use the self-reducibility property to bound certain higher-dimensional
exponential sums which originate from the theory of quantum chaos, thereby
obtaining a proof of one of the main statements in the field---the Hecke quantum unique ergodicity theorem for a generic linear symplectomorphism of
the $2N$-dimensional torus.

\subsection{Quantum chaos problem}
One of the main motivational problems in quantum chaos is \cite{B1, B2, M, S}
describing eigenstates
\begin{equation*}
\widetilde{H}\Psi =\lambda \Psi, \,\,\,\Psi \in \mathcal{H},
\end{equation*}%
of a chaotic Hamiltonian%
\begin{equation*}
\widetilde{H}=Op(H):\mathcal{H\rightarrow H},
\end{equation*}%
where $\mathcal{H}$ is a Hilbert space. We deliberately use the notation $%
Op(H)$ to emphasize the fact that the quantum Hamiltonian $\widetilde{H}$ is
a quantization of a classical Hamiltonian $H:M\rightarrow
\mathbb{C},$
where $M$ is a classical symplectic phase space (usually the cotangent bundle of a
configuration space $M=T^{\ast }X$, in which case $\mathcal{H=}L^{2}(X)$).
In general, describing $\Psi $ is considered
to be an extremely complicated problem. Nevertheless, for a few mathematical
models of quantum mechanics rigorous results have been obtained. We shall
proceed to describe one of these models.

\subsubsection{Hannay--Berry model}
In \cite{HB} Hannay and Berry explored a model for
quantum mechanics on the two-dimensional symplectic torus $(\mathbb{T}%
,\omega )$. Hannay and Berry suggested to quantize simultaneously the
functions on the torus and the linear symplectic group $\Gamma \simeq SL_{2}(%
\mathbb{Z}
)$. One of their main motivations was to study the phenomenon of quantum
chaos in this model \cite{M, R2}. More precisely, they considered an ergodic
discrete dynamical system on the torus which is generated by a hyperbolic
automorphism $A\in \Gamma $. Quantizing the system, the classical phase
space $(\mathbb{T},\omega )$ is replaced by a \textit{finite dimensional}
Hilbert space $\mathcal{H}$, classical observables, i.e., functions $f\in
C^{\infty }(\mathbb{T)}$, by operators $\pi (f)\in \textrm{End}(\mathcal{H)}$, and
classical symmetries by a unitary representation $\rho :\Gamma \rightarrow U(%
\mathcal{H)}$.

\subsubsection{Shnirelman's theorem}
Analogous with the case of the Schr\"{o}dinger equation, consider the
following eigenstates problem
\begin{equation*}
\rho (A)\Psi =\lambda \Psi .
\end{equation*}
A fundamental result, valid for a wide class of quantum systems which are
associated to ergodic classical dynamics, is Shnirelman's theorem \cite{Sh},
asserting that in the semi-classical limit almost all (in a suitable sense) eigenstates become
equidistributed in an appropriate sense.

A variant of Shnirelman's theorem also holds in our situation \cite{BD}.
More precisely, we have that in the semi-classical limit $\hslash
\rightarrow 0$ for almost all (in a suitable sense) eigenstates $\Psi $ of the operator $\rho (A)
$ the corresponding \textit{Wigner distribution} $\left \langle \Psi |\pi
(\cdot )\Psi \right \rangle :C^{\infty }(\mathbb{T})\rightarrow
\mathbb{C}
$ approaches the phase space average $\int_{\mathbb{T}}\cdot |\omega |$.
In this respect, it seems natural to ask whether there exist exceptional
sequences of eigenstates? Namely, eigenstates that do not obey the
Shnirelman's rule (\textit{scarred} eigenstates). It was predicted by Berry \cite%
{B1, B2} that \textit{scarring} phenomenon is not expected to be seen for quantum
systems associated with \textit{generic} chaotic classical dynamics. However, in
our situation the operator $\rho (A)$ is not generic, and exceptional
eigenstates were constructed. Indeed, it was confirmed mathematically in
\cite{FND} that certain $\rho (A)$-eigenstates might localize. For example,
in that paper a sequence of eigenstates $\Psi $ was constructed, for which
the corresponding Wigner distribution approaches the measure $%
{\frac12}%
\delta _{0}+%
{\frac12}%
|\omega |$ on $\mathbb{T}$.

\subsubsection{Hecke quantum unique ergodicity}
A quantum system that obeys Shnirelman's rule is also called quantum
ergodic. Can one impose some natural conditions on the eigenstates so that
no exceptional eigenstates will appear? Namely, \textit{quantum unique
ergodicity} will hold. This question was addressed in a paper by Kurlberg
and Rudnick \cite{KR}. In that paper, they formulated a rigorous notion of
Hecke quantum unique ergodicity for the case $\hslash =1/p$. The following
is a brief description of that work. The basic observation is that the
degeneracies of the operator $\rho (A)$ are coupled with the existence of
symmetries. There exists a commutative group of operators that commutes with
$\rho (A)$, which can in fact be computed. In more detail, the
representation $\rho $ factors through the quotient group $Sp=SL_{2}(\mathbb{%
F}_{p})$. We denote by $T_{A}\subset Sp$ the centralizer of the element $A$,
now considered as an element of the quotient group. The group $T_{A}$ is
called (cf. \cite{KR}) the \textit{Hecke torus} corresponding to the element
$A$. The Hecke torus acts semisimply on $\mathcal{H}$. Therefore, we have a
decomposition

\begin{equation*}
\mathcal{H=}\bigoplus \limits_{\chi :T_{A}\rightarrow
\mathbb{C}
^{\times }}\mathcal{H}_{\chi },
\end{equation*}%
where $\mathcal{H}_{\chi }$ is the Hecke eigenspace corresponding to the
character $\chi $. Consider a unit eigenstate $\Psi \in \mathcal{H}_{\chi }$
and the corresponding Wigner distribution $\mathcal{W}_{\chi }:C^{\infty }(%
\mathbb{T})\mathbb{\rightarrow
\mathbb{C}
}$, defined by the formula $\mathcal{W}_{\chi }(f)=\left \langle \Psi |\pi
(f)\Psi \right \rangle .$ The main statement in \cite{KR} proves an explicit
bound on the semi-classical asymptotic of $\mathcal{W}_{\chi }(f)$

\begin{equation*}
\left \vert \mathcal{W}_{\chi }(f)-\int \limits_{\mathbb{T}}f|\omega
|\right \vert \leq \frac{C_{f}}{p^{1/4}},
\end{equation*}%
where $C_{f}$ is a constant that depends only on the function $f$. In
Rudnick's lectures at MSRI, Berkeley 1999 \cite{R1}, and ECM, Barcelona 2000
\cite{R2}, he conjectured that a stronger bound should hold true, i.e.,

\begin{equation}
\left \vert \mathcal{W}_{\chi }(f)-\int \limits_{\mathbb{T}}f|\omega
|\right \vert \leq \frac{C_{f}}{p^{1/2}}.  \label{estimate}
\end{equation}

A particular case (which implies (\ref{estimate})) of the above inequality
is when $f=\xi $, where $\xi $ is a non-trivial character. In this case, the
integral $\int_{\mathbb{T}}\xi |\omega |$ vanishes and in addition it turns
out that $C_{\xi }=2+o(1)$. Hence, we obtain the following simplified form
of (\ref{estimate})
\begin{equation}
\left \vert \mathcal{W}_{\chi }(\xi )\right \vert \leq \frac{2+o(1)}{\sqrt{p}},
\label{simplified}
\end{equation}%
for sufficiently large $p.$ These stronger bounds were proved in the paper
\cite{GH3}. It will be instructive to briefly recall the main ideas and
techniques used in \cite{GH3}.

\subsubsection{Geometric approach}
The basic observation to be made is that the theory of quantum mechanics on
the torus, in the case $\hslash =1/p$, can be equivalently recast in the
language of the representation theory of finite groups in characteristic $p$. We
will endeavor to give a more precise explanation of this matter. Consider
the quotient $\mathbb{F}_{p}$-vector space $V=\mathbb{T}^{\vee }/p\mathbb{T}%
^{\vee }$, where $\mathbb{T}^{\vee }\simeq
\mathbb{Z}
^{2}$ is the lattice of characters on $\mathbb{T}$. We denote by $H$ $=H(V)$
the Heisenberg group associated to $V$. The group $Sp$ is naturally
identified with the group of linear symplectomorphisms of $V$. We have an
action of $Sp$ on $H$. The Stone--von Neumann theorem (see Theorem \ref{SVN})
states that there exists a unique irreducible representation $\pi
:H\rightarrow GL(\mathcal{H)}$, with a non-trivial character $\psi $ of the
center of $H$. As a consequence of its uniqueness, its isomorphism class is
fixed by $Sp$. This is equivalent to saying that $\mathcal{H}$ is equipped
with a compatible projective representation $\rho:Sp\rightarrow PGL(%
\mathcal{H)}$, which in fact can be linearized to an honest
representation. This representation is the celebrated Weil representation.
Noting that $Sp$ is the group of rational points of the algebraic group $%
\mathbf{Sp}$ (we use boldface letters to denote algebraic varieties), it is
natural to ask whether there exists an algebro-geometric object that
underlies the representation $\rho $. The answer to this question is
positive. The construction is proposed in an unpublished letter of Deligne
to Kazhdan \cite{D}, which appears now in \cite{GH3, GH6}. Briefly, the
content of this letter is a construction of \textit{representation sheaf} $%
\mathcal{K}_{\rho }$ on the algebraic variety $\mathbf{Sp}$. We obtain, as
a consequence, the following general principle:\smallskip

\begin{description}
\item[\textbf{Motivic principle.}] \textit{All quantum mechanical quantities in the
Hannay--Berry model are motivic in nature}\emph{.}
\end{description}

By this we mean that every quantum-mechanical quantity $\mathcal{Q}$ is
associated with a vector space $V_{\mathcal{Q}}$ (certain cohomology of a suitable $\ell $-adic sheaf) endowed with a Frobenius action $\textrm{Fr}:V_{\mathcal{Q}}\rightarrow V_{\mathcal{Q}}$ so that $\mathcal{Q=}%
\textrm{Tr}(\textrm{Fr}_{|V_{\mathcal{Q}}}).$ In particular, it was shown in \cite{GH3} that
there exists a two-dimensional vector space $V_{\chi }$, endowed
with an action $\textrm{Fr}:V_{\chi }\rightarrow V_{\chi }$, so that
\begin{equation}
\mathcal{W}_{\chi }(\xi )=\textrm{Tr}(\textrm{Fr}_{|V_{\chi }}).  \label{motivic}
\end{equation}

This, combined with the purity condition that the eigenvalues of $\textrm{Fr}$ are of
absolute value $1/\sqrt{p},$ implies the estimate (\ref{simplified}).

\subsubsection{The higher-dimensional Hannay--Berry model}
The higher-dimensional Hannay--Berry model is obtained as a quantization of a
$2N$-dimensional symplectic torus $(\mathbb{T},\mathbb{\omega )}$ acted upon
by the group $\Gamma \simeq Sp(2N,%
\mathbb{Z}
)$ of linear symplectic automorphisms. It was first constructed in \cite{GH2}%
, where, in particular, a quantization of the whole group of symmetries $%
\Gamma $ was obtained. Consider a regular ergodic element $A\in \Gamma $,
i.e., $A$ generates an ergodic discrete dynamical system and it is regular
in the sense that it has distinct eigenvalues over $%
\mathbb{C}
.$ It is natural to ask whether quantum unique ergodicity will hold true in
this setting as well, as long as one takes into account the whole group of
hidden (Hecke) symmetries? Interestingly, the answer to this question is NO!
Several new results in this direction have been announced recently. In the
case where the automorphism $A$ is \textit{non-generic}, meaning that it has
an invariant Lagrangian (and more generally co-isotropic) sub-torus $\mathbb{%
T}_{L}\subset \mathbb{T}$, an interesting new phenomenon was revealed. There
exists a sequence $\left \{ \Psi _{\hslash }\right \} $\textit{\ } of Hecke
eigenstates which are closely related to the physical phenomena of \textit{localization}, known in the physics literature (cf. \cite{He, KH}) as \textit{scars}. We will call them \textit{Hecke scars}. These states are localized in the sense
that the associated Wigner distribution converges to the Haar measure $\mu $ on the invariant Lagrangian sub-torus
\begin{equation}
\mathcal{W}_{\Psi _{\hslash }}(f)\rightarrow \int \limits_{\mathbb{T}%
_{L}}f d\mu ,\text{ as }\hslash \rightarrow 0,  \label{localization}
\end{equation}%
for every smooth observable $f$. These special kinds of Hecke eigenstates
were first established in \cite{G}. The semi-classical interpretation of the
localization phenomena (\ref{localization}) was announced in \cite{Ke}.

The above phenomenon motivates the following definition:

\begin{definition}
\label{qergodic}An element $A\in \Gamma $ is called {\normalsize} \textit{generic} if it is regular and admits no non-trivial invariant
co-isotropic sub-tori.
\end{definition}

\begin{remark}
The collection of generic elements constitutes an open subscheme of $\
\Gamma$. In particular, a generic element need not be ergodic automorphism of $
\mathbb{T}$. However, in the case where $\Gamma \simeq SL_{2}(%
\mathbb{Z}
)$ every ergodic (i.e., hyperbolic) element is generic. An example of a
generic element which is not ergodic is given by the Weyl element
$\mathrm{w=}
\left(
\begin{array}{cc}
0 & 1 \\
-1 & 0
\end{array}\right)$.
\end{remark}

In these notes we will require the automorphism $A\in \Gamma $ to be
generic. This case was first considered in \cite{GH4}, where using similar
geometric techniques as in \cite{GH3} the analogue of inequality (\ref{simplified})
was obtained. For the sake of simplicity, let us assume
that the automorphism $A$ is \textit{strongly generic}, i.e., it has no
non-trivial invariant sub-tori.

\begin{theorem}[\cite{GH4}]
Let $\xi $ be a non-trivial character of $\mathbb{T}$. The following bound
holds
\begin{equation}
\left \vert \mathcal{W}_{\chi }(\xi )\right \vert \leq \frac{\lbrack
2+o(1)]^{N}}{\sqrt{p}^{N}},  \label{b}
\end{equation}%
where $p$ is a sufficiently large prime number.\smallskip
\end{theorem}
In particular, using the bound (\ref{b}), we obtain the following statement
for general observable:
\begin{corollary}[Hecke quantum unique ergodicity]
Consider an observable $f$ $\in C^{\infty }(\mathbb{T)}$ and a sufficiently
large prime number $p.$ Then
\begin{equation*}
\left \vert \mathcal{W}_{\chi }(f)-\int \limits_{\mathbb{T}}fd\mu \right \vert
\leq \frac{C_{f}}{\sqrt{p}^{N}},
\end{equation*}%
where $\mu =|\omega |^{N}$ is the corresponding volume form and $C_{f}$ is
an explicit computable constant which depends only on the function $f.$
\end{corollary}

In these notes, using the self-reducibility property of the Weil
representation, we improve the above estimates and obtain the following
theorem:
\begin{theorem}[Sharp bound]
\textbf{\label{rate theorem}}Let $\xi $ be a non-trivial character of $\mathbb{T}$. For sufficiently large prime number $p$ we have
\begin{equation}
\left \vert \mathcal{W}_{\chi }(\xi )\right \vert \leq \frac{\lbrack
2+o(1)]^{r_{p}}}{\sqrt{p}^{N}},  \label{simplified3}
\end{equation}
where the number $r_{p}$ is an integer between $1$ and $N$, that we will call the
\textit{symplectic rank} of $T_{A}$.
\end{theorem}

\begin{remark}
It will be shown $\left( \text{see Subsection \ref{St}}\right) $ that the
distribution of the symplectic rank $r_{p\text{ }}($\ref{simplified3}$)$ in
the set $\left \{ 1,...,N\right \} $ is governed by the Chebotarev density
theorem applied to a suitable Galois group. For example, in the case where
$A\in Sp(4,%
\mathbb{Z}
)$ is strongly generic we have
\begin{equation*}
\lim_{x\rightarrow \infty }\frac{\# \{r_{p}=r\text{ }|\text{ }p\leq x\}}{\pi
(x)}=\tfrac{1}{2},\text{ \  \  \  \  \  \ }r=1,\text{ }2,
\end{equation*}%
where $\pi (x)$ denotes the number of primes up to $x$.
\end{remark}

\begin{remark}
For the more general version of Theorem \ref{rate theorem}, one that holds
in the general generic case $($Definition \ref{qergodic}$)$, see Subsection %
\ref{GGC}.
\end{remark}

In order to witness the improvement of (\ref{simplified3}) over (\ref{b}),
it would be instructive to consider the following extreme scenario. Assume
that the Hecke torus $T_{A}$ acts on $V\simeq \mathbb{F}_{p}^{2N}$
irreducibly. In this case it turns out that $r_{p}=1$. Hence, (\ref%
{simplified3}) becomes%
\begin{equation*}
\left \vert \mathcal{W}_{\chi }(\xi )\right \vert \leq \frac{2+o(1)}{\sqrt{p}%
^{N}},
\end{equation*}%
which constitutes a significant improvement over the coarse topological
estimate (\ref{b}). Let us elaborate on this. Recall the motivic
interpretation (\ref{motivic}) of the Wigner distribution. In \cite{GH4} an
analogous interpretation was given to the higher-dimensional Wigner
distributions, realizing them as $\mathcal{W}_{\chi }(\xi )=\textrm{Tr}(\textrm{Fr}_{|V_{\chi
}}),$ where, by the purity condition, the eigenvalues of $\textrm{Fr}$ are of
absolute value $1/\sqrt{p}^{N}$. But, in this setting the dimension of $%
V_{\chi }$ is not $2$, but $2^{N}$, i.e., the Frobenius looks like
\begin{equation*}
\textrm{Fr}=%
\begin{pmatrix}
\lambda _{1} & \ast  & \ast  & \ast  \\
& \cdot  & \ast  & \ast  \\
&  & \cdot  & \ast  \\
&  &  & \lambda _{2^{N}}%
\end{pmatrix}.%
\end{equation*}%
Hence, if we use only this amount of information, then the best
estimate which can be obtained is (\ref{b}). Therefore, in this respect the
problem that we confront is showing cancellations between different
eigenvalues, more precisely angles, of the Frobenius operator acting on a
high-dimensional vector space, i.e., cancellations in the sum $%
\sum_{j=1}^{2^{N}}e^{i\theta _{j}},$ where the angles $0\leq \theta
_{j}<2\pi $ are defined via $\lambda _{j}=e^{i\theta _{j}}/\sqrt{p}^{N}.$
This problem is of a completely different nature, which is not accounted for
by standard cohomological techniques (we thank R. Heath-Brown for pointing out to us \cite{H} about the phenomenon
of cancellations between Frobenius eigenvalues in the presence of
high-dimensional cohomologies).

\begin{remark}
Choosing a realization $\mathcal{H}\simeq \mathbb{C}(\mathbb{F}_{p}^{N})$, the matrix
coefficient $\mathcal{W}_{\chi }(\xi )$ is equivalent to an exponential sum
of the form
\begin{equation}
\left \langle \Psi |\pi (\xi )\Psi \right \rangle =\sum \limits_{x\in \mathbb{F%
}_{p}^{N}}\Psi (x)e^{\frac{2\pi i}{p}\xi _{+}x}\overline{\Psi }(x+\xi _{-}).
\label{sum}
\end{equation}%
Here one encounters two problems. First, it is not so easy to describe the
eigenstates $\Psi $. Second, the sum $\left( \text{\ref{sum}}\right) $ is a
high-dimensional exponential sum (over $\mathbb{F}_{p})$, which is known to
be hard to analyze using standard techniques. The crucial point that we
explain in these notes is that it can be realized, essentially, as
a one-dimensional exponential sum over $\mathbb{F}_{q},$ where $q=p^{N}$.
\end{remark}

\subsection{Solution via self-reducibility\label{S}}
Let us explain the main idea underlying the proof of estimate (\ref%
{simplified3}). Let us assume for the sake of simplicity that the Hecke
torus is completely inert, i.e., acts irreducibly on the vector space $%
V\simeq \mathbb{F}_{p}^{2N}.$

\subsubsection{Representation theoretic interpretation of the Wigner
distribution}
The Hecke eigenstate $\Psi $ is a vector in a representation space $\mathcal{%
H}$. The space $\mathcal{H}$ supports the Weil representation of the
symplectic group $Sp\simeq Sp(2N,k)$, $k=\mathbb{F}_{p}.$ The vector $\Psi $
is completely characterized in representation theoretic terms, as being a
character vector of the Hecke torus $T_{A}$. As a consequence, all
quantities associated to $\Psi $, and in particular the Wigner distribution $%
\mathcal{W}_{\chi }$ are characterized in terms of the Weil representation.
The main observation to be made is that the Hecke state $\Psi $ can be
characterized in terms of another Weil representation, this time of a group
of much smaller dimension. In fact, it can be characterized, roughly, in
terms of the Weil representation of $SL_{2}(K),$ $K=\mathbb{F}_{p^{N}}$.

\subsubsection{Self-reducibility property}
A fundamental notion in our study is that of a \textit{symplectic module
structure}. A symplectic module structure is a triple $(K,V,\overline{\omega})$,
where $K$ is a finite dimensional commutative algebra over $k,$ equipped
with an action on the vector space $V$, and $\overline{\omega }$ is a $K$%
-linear symplectic form satisfying the property $\textrm{Tr}_{K/k}(\overline{\omega }%
)=\omega .$ Let us assume for the sake of simplicity that $K$ is a field.
Let $\overline{Sp}=Sp(V,\overline{\omega })$ be the group of $K$-linear
symplectomorphisms with respect to the form $\overline{\omega }$. There
exists a canonical embedding
\begin{equation}
\iota :\overline{Sp}\hookrightarrow Sp.  \label{CI}
\end{equation}

We will be mainly concerned with symplectic module structures which are
associated to maximal tori in $Sp$. More precisely, it will be shown that
associated to a maximal torus $T\subset Sp$ there exists a canonical
symplectic module structure $(K,V,\overline{\omega })$ so that $T\subset
\overline{Sp}$. The most extreme situation is when the torus $T\subset Sp$
is completely inert, i.e., acts irreducibly on the vector space $V$. In this
particular case, the algebra $K$ is in fact a field with $\dim _{K}V=2$
which implies that $\overline{Sp}\simeq SL_{2}(K)$, i.e., using (\ref{CI})
we get $T\subset SL_{2}(K)\subset Sp.$

Let us denote by $(\rho ,Sp,\mathcal{H})$ the Weil representation of $Sp$.
The main observation now is (cf. \cite{Ge}) the following:

\begin{theorem}[Self-reducibility property]\label{SRD} The restricted representation $(\overline{\rho }=\iota ^{\ast
}\rho ,SL_{2}(K),\mathcal{H})$ is the Weil representation of $SL_{2}(K)$.
\end{theorem}

Applying the self-reducibility property to the Hecke torus $T_{A}$, it
follows that the Hecke eigenstates $\Psi $ can be characterized in terms of
the Weil representation of $SL_{2}(K)$. Therefore, in this respect, Theorem %
\ref{rate theorem} is reduced to the result obtained in \cite{GH3}.

\subsection{Quantum unique ergodicity for statistical states}
Let $A\in \Gamma $ be a generic linear symplectomorphism. As in harmonic
analysis, one would like to use Theorem \ref{rate theorem} concerning the
Hecke eigenstates in order to extract information on the spectral theory of
the operator $\rho (A)$ itself. For the sake of simplicity, let us assume
that $A$ is strongly generic, i.e., it acts on the torus $\mathbb{T}$ with
no non-trivial invariant sub-tori. Next, a possible reformulation of the
quantum unique ergodicity statement, one which is formulated for the
automorphism $A$ itself instead of the Hecke group of symmetries, is
presented.

The element $A$ acts via the Weil representation $\rho $ on the space $%
\mathcal{H}$ and decomposes it into a direct sum of $\rho \left( A\right) $%
-eigenspaces
\begin{equation}
\mathcal{H=}\bigoplus \limits_{\lambda \in Spec(\rho \left( A\right) )}%
\mathcal{H}_{\lambda }.  \label{dsA}
\end{equation}

Considering an $\rho \left( A\right) $-eigenstate $\Psi $ and the
corresponding projector $P_{\Psi }$ one usually studies the Wigner
distribution $\left \langle \Psi |\pi (\xi )\Psi \right \rangle =\textrm{Tr}(\pi (\xi
)P_{\Psi })$ which, due to the fact that $rank(P_{\Psi })=1,$ is sometimes
called$\ $\textit{pure state.} In the same way, we might think about an
Hecke--Wigner distribution $\left \langle \Psi |\pi (\xi )\Psi \right \rangle
=\textrm{Tr}(\pi (\xi )P_{\chi }),$ attached to a $T_{A}$-eigenstate $\Psi $, as a
\textit{pure Hecke state}. Following von Neumann \cite{vN} we suggest the
possibility of looking at the more general \textit{statistical state},
defined by a non-negative, self-adjoint operator $D$, called the von Neumann
density operator, normalized to have $\textrm{Tr}(D)=1.$ For example, to the
automorphism $A$ we can attach the natural family of density operators $%
D_{\lambda }=\frac{1}{m_{\lambda }}P_{\lambda ,}$ where $P_{\lambda }$ is
the projector on the eigenspace $\mathcal{H}_{\lambda }$ $\left( \text{\ref%
{dsA}}\right) ,$ and $m_{\lambda }=\dim (\mathcal{H}_{\lambda }).$
Consequently, we obtain a family of statistical states
\begin{equation*}
\mathcal{W}_{\lambda }\left( \cdot \right) =\textrm{Tr}(\pi (\cdot )D_{\lambda }).
\end{equation*}

\begin{theorem}
\label{QUESS} Let $\xi $ be a non-trivial character of $\mathbb{T}$. For a
sufficiently large prime number $p$, and every statistical state $\mathcal{W}%
_{\lambda },$ we have
\begin{equation}
|\mathcal{W}_{\lambda }(\xi )|\leq \frac{(2+o(1))^{r_{p}}}{\sqrt{p}^{N}},
\label{BQUESS}
\end{equation}%
where $r_{p}$ is an explicit integer $1\leq r_{p}\leq N$ which is determined by
$A.$
\end{theorem}

Theorem \ref{QUESS} follows from the fact that the Hecke torus $T_{A}$ acts
on the spaces $\mathcal{H}_{\lambda },$ and hence, one can use the Hecke
eigenstates, and the bound (\ref{simplified3}).
In particular, using (\ref{BQUESS}) we obtain for a general observable the following bound:
\begin{corollary}
Consider an observable $f$ $\in C^{\infty }(\mathbb{T)}$ and a sufficiently
large prime number $p.$ Then
\begin{equation*}
\left \vert \mathcal{W}_{\lambda }(f)-\int \limits_{\mathbb{T}}fd\mu
\right \vert \leq \frac{C_{f}}{\sqrt{p}^{N}},
\end{equation*}%
where $\mu =|\omega |^{N}$ is the corresponding volume form and $C_{f}$ is
an explicit computable constant which depends only on the function $f.$
\end{corollary}

\subsection{Results}
\begin{enumerate}
\item \emph{Bounds\ of higher-dimensional exponential sums. }The main
results of these notes are a sharp estimates of certain higher-dimensional
exponential sums attached to tori in $Sp(2N,\mathbb{F}_q)$. This is the content of
Theorems \ref{KR1} and \ref{KR_general1} and is obtained using the self-reducibility
property of the Weil representation as stated in Theorems \ref{reWeil} and \ref{Weildecomp}.

\item \emph{Hecke quantum unique ergodicity theorem. }The main application
of these notes is the proof of the \textit{Hecke quantum unique ergodicity
theorem}, i.e., Theorems \ref{SGC} and \ref{GCT}, for generic linear
symplectomorphism of the torus in any dimension. The proof of the theorem is
a direct application of the sharp bound on the higher-dimensional
exponential sums.

\item \emph{Multiplicities formula.} Exact formula for the multiplicities,
i.e., the dimensions of the character spaces for the action of maximal tori
in the Weil representation are derived. This is obtained first for the $%
SL_{2}(\mathbb{F}_{q})$ case in Theorem \ref{Multiplicities} using the
character formula presented in Theorem \ref{ChF}. Then, as a direct
application of the self-reducibility property, the formula is extended in
Theorem \ref{MF} to the higher-dimensional cases.
\end{enumerate}

In addition, a formulation of the quantum unique ergodicity statement for
quantum chaos problems, close in spirit to the von Neumann idea about
density operator, is suggested in Theorem \ref{QUESS}. The statement
includes only the quantum operator $A$ rather than the whole Hecke group of
symmetries \cite{KR}. The proof of the statement uses the Hecke operators as
a harmonic analysis tool.

\subsection{Structure of the notes}
Apart from the introduction, the notes consist of five sections.

In Section \ref{PLS} we give some preliminaries on representation theory
which are used in the notes. In Subsection \ref{IP} we recall the invariant
presentation of the Weil representation over finite fields \cite{GH6}, and
we discuss applications to multiplicities. Section \ref{selfreducibility}
constitutes the main technical part of this work. Here we develop the theory
that underlies the self-reducibility property of the Weil representation. In
particular, in Subsection \ref{Sms} we introduce the notion of symplectic
module structure. In Subsection \ref{SmsT} we prove the existence of
symplectic module structure associated with a maximal torus in $Sp$.
Finally, we establish the self-reducibility property of the Weil
representation, i.e., Theorem \ref{Weildecomp}, and apply this property to
get information on multiplicities in Subsection \ref{HDM}. Section \ref{B}
is devoted to an application of the theory developed in previous sections to
estimating higher-dimensional exponential sums which originate from the mathematical
theory of quantum chaos. In Section \ref{HBM} we describe the higher-dimensional
Hannay--Berry model of quantum mechanics on the torus. Finally, in Section \ref{HQUE}
we present the main application of these notes---the proof of the Hecke quantum unique ergodicity theorem for
generic linear symplectomorphisms of the $2N$-dimensional torus.

\begin{remark}
Complete proofs for the statements appearing in these notes will be given
elsewhere.
\end{remark}

\subsection{Acknowledgments}
It is a pleasure to thank our teacher J. Bernstein for his interest and
guidance in this project. We thank D. Kazhdan for several interesting
discussions. We appreciate the support of the Technion, and in particular
of M. Baruch. We thank D. Kelmer for sharing with us computer simulation
data. Finally, we would like to thank O. Ceyhan and the organizers of the
conference AGAQ, held in Istanbul during June 2006, and J. Wolf and the
organizers of the conference Lie Groups, Lie Algebras and Their
Representations, Berkeley, November 2006, for the invitation to present
this work.
\section{Preliminaries}\label{PLS}

In this section, we denote by $k=\mathbb{F}_{q}$ the finite field of $q$
elements and odd characteristic.

\subsection{The Heisenberg representation}

Let $(V,\omega )$ be a $2N$-dimensional symplectic vector space over the
finite field $k$. There exists a two-step nilpotent group $H=H\left(
V,\omega \right) $ associated to the symplectic vector space $(V,\omega )$.
\ The group $H$ is called the \textit{Heisenberg group.} It can be realized
as the set $H=V\times k,$ equipped with the multiplication rule
\begin{equation*}
(v,z)\cdot (v^{\prime },z^{\prime })=(v+v^{\prime },z+z^{\prime }+\tfrac{1}{2%
}\omega (v,v^{\prime })).
\end{equation*}

The center of $H$ is $Z(H)=\{(0,z):z\in k\}$. $\ $Fix a non-trivial central
character $\psi :Z(H)\longrightarrow
\mathbb{C}
^{\times }$. We have the following fundamental theorem:

\begin{theorem}[Stone--von Neumann]
\label{SVN}
There exists a unique $($up to isomorphism$)$ irreducible representation $(\pi ,H,\mathcal{H})$ with central character $\psi $, i.e., $\pi (z)=\psi(z)\textrm{Id}_{\mathcal{H}}$ for every $z\in Z(H).$
\end{theorem}

We call the representation $\pi $ appearing in Theorem \ref{SVN}, the
\textit{Heisenberg representation} associated with the central character $%
\psi $.

\begin{remark}
\label{Schrodinger}The representation $\pi $, although it is unique, admits
a multitude of different models $($realizations$)$. In fact, this is one of
its most interesting and powerful attributes. In particular, to any
Lagrangian splitting $V=L^{\prime }\oplus L$, there exists a model $(\pi
_{L^{\prime },L},H,%
\mathbb{C}
(L))$, where $%
\mathbb{C}
(L)$  denotes the space of complex valued functions on $L$. In this model,
we have the following actions:

\begin{itemize}
\item $\pi _{L^{\prime },L}(l^{\prime })[f](x)=\psi (\omega (l^{\prime
},x))f(x)$;

\item $\pi _{L^{\prime },L}(l)[f](x)=f(x+l)$;

\item $\pi _{L^{\prime },L}(z)[f](x)=\psi (z)f(x)$,

where $l^{\prime }\in L^{\prime }$, $x,l\in L$, and $z\in Z(H).$
\end{itemize}
\end{remark}

The above model is called the Schr\"{o}dinger realization associated
with the splitting $V=L^{\prime }\oplus L$.

\subsection{The Weyl transform\label{Weyl_sub}}

Given a linear operator $A:\mathcal{H\rightarrow H}$ we can associate to it
a function on the group $H$ defined as follows%
\begin{equation}
W(A)(h)=\tfrac{1}{\dim \mathcal{H}}\textrm{Tr}(A\pi (h^{-1})).  \label{WT}
\end{equation}%
The transform $W:\textrm{End}(\mathcal{H)\rightarrow }%
\mathbb{C}
(H)$ is called the \textit{Weyl transform \cite{H2, We1}.}
The Weyl transform admits a left inverse $\pi:\mathbb{C}(H)\rightarrow \textrm{End}(\mathcal{H})$ given by the extended action $\pi(K)=\sum\limits_{h\in H}K(h)\pi(h)$.

\subsection{The Weil representation \label{WR}}
Let $Sp=Sp(V,\omega )$ denote the group of linear symplectic automorphisms
of $V$. The group $Sp$ acts by group automorphisms on the Heisenberg group
through its tautological action on the vector space $V$. A direct
consequence of Theorem \ref{SVN} is the existence of a projective
representation $\widetilde{\rho }:Sp\rightarrow PGL(\mathcal{H)}$. The
classical construction of $\widetilde{\rho }$ out of the Heisenberg
representation $\pi $ is due to Weil \cite{W}. Considering the Heisenberg
representation $\pi $ and an element $g\in Sp$, one can define a new
representation $\pi ^{g}$ acting on the same Hilbert space via $\pi
^{g}\left( h\right) =\pi \left( g\left( h\right) \right) $. Clearly both $%
\pi$ and $\pi ^{g}$ have central character $\psi$; hence, by Theorem \ref{SVN},
they are isomorphic. Since the space $\mathsf{Hom}_{H}(\pi ,\pi ^{g})$
is one-dimensional, choosing for every $g\in Sp$ a non-zero representative $%
\widetilde{\rho }(g)\in \mathsf{Hom}_{H}(\pi ,\pi ^{g})$ gives the required
projective representation. In more concrete terms, the projective
representation $\widetilde{\rho }$ is characterized by the formula
\begin{equation}
\widetilde{\rho }\left( g\right) \pi \left( h\right) \widetilde{\rho }\left(
g^{-1}\right) =\pi \left( g\left( h\right) \right) ,  \label{Eg}
\end{equation}%
for every $g\in Sp$ and $h\in H$. It is a peculiar phenomenon of the
finite field setting that the projective representation $\widetilde{\rho }$ can be linearized into an honest representation. This linearization is unique, except in the case the finite field is $\mathbb{F}_{3}$ and $\dim V=2
$ (for the canonical choice in the latter case see \cite{GH7}).
\begin{theorem}
\label{linearization} There exists a canonical unitary representation
\begin{equation*}
\rho :Sp\longrightarrow GL(\mathcal{H)},
\end{equation*}
satisfying the formula (\ref{Eg}).
\end{theorem}

\subsubsection{Invariant presentation of the Weil representation\label{IP}}
An elegant description of the Weil representation can be obtained \cite{GH6} using the
Weyl transform (see Subsection \ref{Weyl_sub}). Given an element $g\in Sp$, the operator $\rho (g)
$ can be written as $\rho (g)=\pi (K_{g})$, where $K_{g}$ is the Weyl
transform $K_{g}=W(\rho (g))$. The homomorphism property of $\rho$ is manifested as
$$
K_{g}\ast K_{h}=K_{gh}\text{ \  \ for every \ }g,h\in Sp,
$$
where $\ast$ denotes (properly normalized) group theoretic convolution on $H$.
Finally, the function $K$ can be explicitly described on an appropriate
subset of $Sp$ \cite{GH6}. Let $U\subset Sp$ denote the subset consisting of
all elements $g\in Sp$ such that $g-I$ is invertible. For every $g\in U$
and $v\in V$ we have
\begin{equation}
K_{g}(v)=\nu (g)\psi (\tfrac{1}{4}\omega (\kappa (g)v,v)),
\label{kernel_formula}
\end{equation}%
where $\kappa (g)=\frac{g+I}{g-I}$ is the Cayley transform \cite{H2, We2}, and
$$
\nu (g)=(G/q)^{2N}\sigma (\det(g-I)),
$$
with $\sigma$ the unique quadratic character of the multiplicative group
$\mathbb{F}_{q}^{\times }$, and $G= \sum\limits_{z\in Z(H)} \psi(z^2)$ the quadratic Gauss sum.

\subsection{The Heisenberg--Weil representation}
Let $J$ denote the semi-direct product $J=Sp\ltimes H.$ The group $J$ is
sometimes referred to as the \textit{Jacobi} group. The compatible pair $%
(\pi ,\rho )$ is equivalent to a single representation $\tau
:J\longrightarrow GL(\mathcal{H})$ of the Jacobi group defined by the
formula $\tau (g,h)=\rho (g)\pi (h)$. It is an easy exercise to verify
that the Egorov identity (\ref{Eg}) implies the multiplicativity of the map $%
\tau $.

In these notes, we would like to adopt the name \textit{Heisenberg--Weil
representation} when referring to the representation $\tau $.

\subsection{Character formulas}
The invariant presentation (\ref{WT}) and formula (\ref{kernel_formula}) imply \cite{GH6}
a formula for the character of the $2N$-dimensional Heisenberg--Weil representation over a finite field
(cf. \cite{Ge, H3}).

\begin{theorem}[Character formulas \cite{GH6}]\label{ChF}
The character $\textrm{ch}_{\rho \text{ }}$ of the Weil representation,
when restricted to the subset $U$, is given by
\begin{equation}
\textrm{ch}_{\rho }(g)=\sigma ((-1)^N\det(g-I)),
\label{Char W}
\end{equation}%
and the character $\textrm{ch}_\tau$ of the Heisenberg--Weil
representation, when restricted to the subset $U\times H$, is given by
\end{theorem}

\begin{equation}
\textrm{ch}_{\tau }(g,v,z)=\textrm{ch}_{\rho }(g)\psi (\tfrac{1}{4}\omega (\kappa (g)v,v)+z).
\label{Char W-H}
\end{equation}

\subsection{Application to multiplicities}
We would like to apply the formula (\ref{Char W-H}) to the study of the
multiplicities arising from actions of tori via the Weil representation (cf.
\cite{AM, Ge, Si}). Let us start with the two-dimensional case (see Theorem %
\ref{MF} for the general case). Let $T\subset Sp\simeq SL_{2}(\mathbb{F}_{q})
$ be a maximal torus. The torus $T$ acts semisimply on $\mathcal{H}$,
decomposing it into a direct sum of character spaces $\mathcal{H=}%
\bigoplus \limits_{\chi :T\rightarrow
\mathbb{C}
^{\times }}\mathcal{H}_{\chi }.$

As a consequence of having the explicit formula (\ref{Char W}), we obtain\ a
simple description for the multiplicities $m_{\chi }=\dim $ $\mathcal{H}%
_{\chi }.$ Denote by $\sigma :T\rightarrow
\mathbb{C}
^{\times }$ the unique quadratic character of $T.$

\begin{theorem}[Multiplicities formula]\label{Multiplicities} We have $m_{\chi }=1$ for any character $\chi \neq $ $%
\sigma .$ Moreover, $m_{\sigma }=2$ or $0,$ depending on whether the torus $%
T $ is split or inert, respectively.
\end{theorem}

What about the multiplicities for action of tori in the Weil representation
of higher-dimensional symplectic groups? This problem can be answered (see Theorem \ref{MF}) using the
self-reducibility property of the Weil representation.

\section{Self-reducibility of the Weil representation\label{selfreducibility}}
In this section, unless stated otherwise, the field $k$ is an \textit{%
arbitrary field} of characteristic different from two.

\subsection{Symplectic module structures\label{Sms}}
Let $K$ be a finite-dimensional commutative algebra over the field $k$.
Let $\textrm{Tr}:$ $K\rightarrow k$ be the trace map, associating to an element $x\in $
$K$ the trace of the $k$-linear operator $m_{x}:$ $K\rightarrow $ $K$
obtained by left multiplication by the element $x$. Consider a symplectic
vector space $(V,\omega )$ over $k.$

\begin{definition}
\label{SMS}A symplectic $K$-module structure on $\left( V,\omega \right) $
is an action $K\otimes _{k}V\rightarrow V,$ and a $K$-linear symplectic
form $\overline{\omega }:V\times V\rightarrow $ $K$ such that
\begin{equation}
\textrm{Tr}\circ \overline{\omega }=\omega .  \label{SC}
\end{equation}
\end{definition}

Given a symplectic module structure $(K,V,\overline{\omega })$ on a
symplectic vector space $(V,\omega ),$ we denote by $\overline{Sp}=Sp(V,%
\overline{\omega })$ the group of $K$-linear symplectomorphisms with respect
to the form $\overline{\omega }$. The compatibility condition (\ref{SC})
gives a natural embedding
\begin{equation}
\iota :\overline{Sp}\hookrightarrow Sp.  \label{im}
\end{equation}%

\subsection{Symplectic module structure associated with a maximal torus\label%
{SmsT}}

Let $T\subset Sp$ be a maximal torus.

\subsubsection{A particular case\label{particular}}

In order to simplify the presentation, let us assume first that $T$ acts
irreducibly on the vector space $V$, i.e., there exists no non-trivial $T$-invariant subspaces.
Let $A=Z(T,\textrm{End}(V)),$ be the centralizer of $T$ in the algebra of all linear endomorphisms. Clearly (due to the assumption of irreducibility) $A$ is a division algebra. Moreover, we have

\begin{claim}
\label{Comm} The algebra $A$ is commutative.
\end{claim}

In particular, this claim implies that $A$ is a field extension of $k$.
Let us now describe a special quadratic element in the Galois group $\textrm{Gal}(A/k)$.
Denote by $(\cdot )^{t\text{ }}:\textrm{End}(V)\rightarrow \textrm{End}(V)$ the symplectic transpose
characterized by the property%
\begin{equation*}
\omega (Rv,u)=\omega (v,R^{t}u),
\end{equation*}%
for all $v,u\in V$, and every $R\in \textrm{End}(V)$. It can be easily verified that $%
(\cdot )^{t}$ preserves $A$, leaving the subfield $k$ fixed, hence, it
defines an element $\Theta \in \textrm{Gal}(A/k),$ satisfying $\Theta ^{2}=\textrm{Id}$.
Denote by $K=A^{\Theta }$ the subfield of $A$ consisting of
elements fixed by $\Theta$. We have the following proposition:

\begin{proposition}[Hilbert's Theorem 90]
\label{prop_dim} We have $\dim _{K}V=2.$
\end{proposition}

\begin{corollary}
We have\ $\dim _{K}A=2.$
\end{corollary}

As a corollary, we have the following description of $T$. Denote by $%
N_{A/K}:A\rightarrow K$ the standard norm map.

\begin{corollary}
\label{DT}We have $T=S(A)=\left \{ a\in A:N_{A/K}(a)=1\right \} $
\end{corollary}

The symplectic form $\omega $ can be lifted to a $K$-linear symplectic
form $\overline{\omega }$, which is invariant under the action of the torus $%
T$. This is the content of the following proposition:

\begin{proposition}[Existence of canonical symplectic module structure]
\label{canonicallift} There exists a\ canonical $T$-invariant $K$-linear
symplectic form $\overline{\omega }:V\times V\rightarrow K$ satisfying the
property $\textrm{Tr}\circ \overline{\omega }=\omega .$
\end{proposition}

Concluding, we obtained a $T$-invariant symplectic $K$-module structure on $%
V $.

Let $\overline{Sp}=Sp(V,\overline{\omega })$ denote the group of $K$-linear
symplectomorphisms with respect to the symplectic form $\overline{\omega }$.
We have (\ref{im}) a natural embedding $\overline{Sp}$ $\subset Sp$. The
elements of $T$ commute with the action of $K$, and preserve the symplectic
form $\overline{\omega }$\ (Proposition \ref{canonicallift}); hence, we can
consider $T$ as a subgroup of $\overline{Sp}$. By Proposition \ref%
{prop_dim} we can identify $\overline{Sp}\simeq SL_{2}(K),$ and using (\ref%
{im}) we obtain
\begin{equation}
T\subset SL_{2}(K)\subset Sp.  \label{inr}
\end{equation}%
To conclude we see that $T$ consists of the $K$-rational points of a
maximal torus $\mathbf{T\subset SL}_{2}$ (in this case $T$ consists of the
rational points of an inert torus).

\subsubsection{General case\label{general}}
Here, we drop the assumption that $T$ acts irreducibly on $V$. By the same
argument as before, the algebra $A=Z(T,\textrm{End}(V))$ is
commutative, yet, it may no longer be a field. The symplectic transpose $%
(\cdot )^{t}$ preserves the algebra $A$, and induces an involution $\Theta
:A\rightarrow A.$ Let $K=A^{\Theta }$ be the subalgebra consisting of
elements $a\in A$ fixed by $\Theta $. Following the same argument as in
the proof of Proposition \ref{prop_dim}, we can show that $V$ is a free $K$%
-module of rank $2$. Following the same arguments as in the proof of
Proposition \ref{canonicallift}, we can show that there exists a canonical
symplectic form $\overline{\omega }:V\times V\rightarrow K$, which is $K$%
-linear and invariant under the action of the torus $T$. Concluding,
associated to a maximal torus $T$ there exists a $T$-invariant symplectic $K$%
-module structure%
\begin{equation}
(K,V,\overline{\omega }).  \label{SMSG}
\end{equation}

Denote by $\overline{Sp}=Sp(V,\overline{\omega })$ the group of $K$-linear
symplectomorphisms with respect to the form $\overline{\omega }.$ We have a
natural embedding
\begin{equation}
\iota _{S}:\overline{Sp}\hookrightarrow Sp  \label{SIG}
\end{equation}%
and we can consider $T$ as a subgroup of $\overline{Sp}$. Finally, we have $%
\overline{Sp}\simeq SL_{2}(K),$ and $T$ consists of the $K$-rational points
of a maximal torus $\mathbf{T\subset SL}_{2}$. In particular, the relation (%
\ref{inr}) holds also in this case: $T\subset SL_{2}(K)\subset Sp.$

We shall now proceed to give a finer description of all objects discussed
so far. The main technical result is summarized in the following lemma:

\begin{lemma}[Symplectic decomposition]
\label{technical}We have a canonical decomposition
\begin{equation}
(V,\omega )=\bigoplus \limits_{\alpha \in \Xi }(V_{\alpha },\omega _{\alpha
}),  \label{SR}
\end{equation}%
into $(T,A)$-invariant symplectic subspaces. In addition, we have the
following associated canonical decompositions
\end{lemma}

\begin{enumerate}
\item $T=\prod T_{\alpha },$ where $T_{\alpha }$ consists of elements $t\in
T$ such that $t_{|V_{\beta }}=\textrm{Id}$ for every $\beta \neq \alpha $.

\item $A=\bigoplus A_{\alpha }$, where $A_{\alpha }$ consists of elements $%
a\in A$ such that $a_{|V_{\beta }}=\textrm{Id}$ for every $\beta \neq \alpha $.
Moreover, each sub-algebra $A_{\alpha }$ is preserved under the involution $%
\Theta .$

\item $K=\bigoplus K_{\alpha }$, where $K_{\alpha }=A_{\alpha }^{\Theta }$.
Moreover, $K_{\alpha }$ is a field and $\dim _{K_{\alpha }}V_{\alpha }=2$.

\item $\overline{\omega }=\bigoplus \overline{\omega }_{\alpha }$, where $%
\overline{\omega }_{\alpha }:V_{\alpha }\times V_{\alpha }\rightarrow
K_{\alpha }$ is a $K_{\alpha }$-linear $T_{\alpha }$-invariant symplectic
form satisfying $\textrm{Tr}\circ \overline{\omega }_{\alpha }=\omega _{\alpha }$.
\end{enumerate}

\begin{definition}
\label{type}We will call the set $\Xi $\ $\left( \text{\ref{SR}}\right) $
the symplectic type of $T$ and the number $\left \vert \Xi \right \vert $ the
symplectic rank of $T$.
\end{definition}

Using the results of Lemma \ref{technical}, we have an isomorphism
\begin{equation}
\overline{Sp}\simeq \prod \overline{Sp}_{\alpha },  \label{decomp5}
\end{equation}%
where $\overline{Sp}_{\alpha }=Sp(V_{\alpha },\overline{\omega }_{\alpha })$
denotes the group of $K_{\alpha }$-linear symplectomorphisms with respect to
the form $\overline{\omega }_{\alpha }$. Moreover, for every $\alpha \in
\Xi $ we have $T_{\alpha }\subset \overline{Sp}_{\alpha }$. In particular,
under the identifications $\overline{Sp}_{\alpha }\simeq SL_{2}(K_{\alpha }),
$ there exist the following sequence of inclusions
\begin{equation}
T=\prod T_{\alpha }\subset \prod SL_{2}(K_{\alpha })=SL_{2}(K)\subset Sp,
\label{IOG}
\end{equation}
and for every $\alpha \in \Xi $ the torus $T_{\alpha }$ coincides with the $%
K_{\alpha }$-rational points of a maximal torus $\mathbf{T}_{\alpha }\subset
\mathbf{SL}_{2}$.

\subsection{Self-reducibility of the Weil representation\label{selfred}}
In this subsection we assume that the field $k$ is a finite field of odd characteristic (although, the results continue to hold true also for local fields of characteristic $\neq 2$, i.e., with the appropriate modification, replacing
the group $Sp$ with its double cover $\widetilde{Sp}$).
Let $(\tau ,J,\mathcal{H)}$ be the Heisenberg--Weil representation
associated with a central character $\psi :Z(J)=Z(H)\rightarrow
\mathbb{C}
^{\times }$. Recall that $J=Sp\ltimes H,$ and $\tau $ is obtained as a
semi-direct product, $\tau =\rho \ltimes \pi$, of the Weil representation $%
\rho $ and the Heisenberg representation $\pi $. Let $T\subset Sp$ be a
maximal torus.

\subsubsection{A particular case\label{particular1}}
For clarity of presentation, assume first that $T$ acts irreducibly on $V
$. Using the results of the previous section, there exists a symplectic
module structure $(K,V,\overline{\omega })$ (in this case $K/k$ is a field
extension of degree $[K:k]=N$). The group $\overline{Sp}=Sp(V,\overline{%
\omega })$ is imbedded as a subgroup $\iota _{S}:\overline{Sp}%
\hookrightarrow Sp.$ Our goal is to describe the restriction
\begin{equation}
(\overline{\rho }=\iota _{S}^{\ast }\rho ,\overline{Sp},\mathcal{H)}.
\label{RR}
\end{equation}

Define an auxiliary Heisenberg group
\begin{equation}
\overline{H}=V\times K,  \label{auxH}
\end{equation}%
with the multiplication given by $(v,z)\cdot (v^{\prime },z^{\prime
})=(v+v^{\prime },z+z^{\prime }+%
{\frac12}%
\overline{\omega }(v,v^{\prime }))$. There exists homomorphism%
\begin{equation}
\iota _{H}:\overline{H}\rightarrow H,  \label{homauxH}
\end{equation}%
given by $(v,z)\mapsto (v,\textrm{Tr}(z))$. Consider the pullback $(\overline{\pi }%
=\iota _{H}^{\ast }\pi ,\overline{H},\mathcal{H)}.$ We have

\begin{proposition}
\label{resHeisenberg} The representation $(\overline{\pi }=$ $\iota
_{H}^{\ast }\pi ,\overline{H},\mathcal{H)}$ is the Heisenberg representation
associated with the central character $\overline{\psi }=\psi \circ \textrm{Tr}$.
\end{proposition}

The group $\overline{Sp}$ acts by automorphisms on the group $\overline{H}$
through its tautological action on the $V$-coordinate. This action is
compatible with the action of $Sp$ on $H$, i.e., we have $\iota _{H}(g\cdot
h)=$ $\iota _{S}(g)\cdot $ $\iota _{H}(h)$ for every $g\in \overline{Sp}$,
and $h\in \overline{H}$. The description of the representation $\overline{%
\rho }$ (\ref{RR}) now follows easily (cf. \cite{Ge}).

\begin{theorem}[Self-reducibility property (particular case)]
\label{reWeil} The representation $(\overline{\rho },\overline{Sp},\mathcal{%
H)}$ is the Weil representation associated with the Heisenberg
representation $(\overline{\pi },\overline{H},\mathcal{H)}$.
\end{theorem}

\begin{remark}
We can summarize the result in a slightly more elegant manner using the
Jacobi groups. Let $J=Sp\ltimes H$ and $\overline{J}=\overline{Sp}\ltimes
\overline{H}$ be the Jacobi groups associated with the symplectic spaces $%
(V,\omega )$ and $(V,\overline{\omega })$ respectively. We have a homomorphism $\iota :\overline{J}\rightarrow J,$ given by $\iota (g,h)=($ $%
\iota _{S}(g),$ $\iota _{H}(h))$. Let $(\tau ,J,\mathcal{H)}$ be the
Heisenberg--Weil representation of $J$ associated with a character $\psi $ of
the center $Z(J)$ (note that $Z(J)=Z(H)$), then the pullback $(\iota ^{\ast
}\tau ,\overline{J},\mathcal{H)}$ is the Heisenberg--Weil representation of $%
\overline{J}$, associated with the character $\overline{\psi }=\psi \circ \textrm{Tr}$
of the center $Z(\overline{J})$.
\end{remark}

\subsubsection{The general case\label{general1}}

Here, we drop the assumption that $T$ acts irreducibly on $V$. Let $(K,V,%
\overline{\omega })$ be the associated symplectic module structure (\ref%
{SMSG}). Using the results of Subsection \ref{general}, we have
decompositions
\begin{eqnarray}
(V,\omega ) =\bigoplus \limits_{\alpha \in \Xi }(V_{\alpha },\omega
_{\alpha }),  \ \ \ \ \
(V,\overline{\omega }) =\bigoplus \limits_{\alpha \in \Xi }(V_{\alpha },%
\overline{\omega }_{\alpha }),  \label{decomm7}
\end{eqnarray}%
where $\overline{\omega }_{\alpha }:V_{\alpha }\times V_{\alpha }\rightarrow
K_{\alpha }$. Let (cf. \ref{auxH}) $\overline{H}=V\times K,$ be the
Heisenberg group associated with $(V,\overline{\omega })$. There exists (cf.
(\ref{homauxH}) a homomorphism $\iota _{H}:\overline{H}\rightarrow H.$ Let
us describe the pullback $\overline{\pi }=\iota _{H}^{\ast }\pi$ of the
Heisenberg representation. First, we note that the decomposition (\ref%
{decomm7}) induces a corresponding decomposition of the Heisenberg group, $%
\overline{H}=\prod \overline{H}_{\alpha },$ where $\overline{H}_{\alpha }$
is the Heisenberg group associated with $(V_{\alpha },\overline{\omega }%
_{\alpha })$. We have the following proposition

\begin{proposition}\label{Heisenbergdecom}
There exists an isomorphism
\begin{equation*}
(\overline{\pi },\overline{H},\mathcal{H)\simeq (}\bigotimes \overline{\pi }%
_{\alpha },\prod \overline{H}_{\alpha },\bigotimes \mathcal{H}_{\alpha }),
\end{equation*}%
where $(\overline{\pi }_{\alpha },\overline{H}_{\alpha },\mathcal{H}_{\alpha
})$ is the Heisenberg representation of $\overline{H}_{\alpha }$ associated
with the central character $\overline{\psi }_{\alpha }=\psi \circ
\textrm{Tr}_{K_{\alpha }/k}$.
\end{proposition}

Let $\iota _{S}:\overline{Sp}\hookrightarrow Sp,$ be the embedding (\ref{SIG}%
). Our next goal is to describe the restriction $\overline{\rho }=\iota
_{S}^{\ast }\rho .$ Recall that we have a decomposition $\overline{Sp}%
=\prod \overline{Sp}_{\alpha }$ (see (\ref{decomp5})). In terms of this
decomposition we have (cf. \cite{Ge})

\begin{theorem}[Self-reducibility property---general case]
\label{Weildecomp} There exists an isomorphism
\begin{equation*}
(\overline{\rho },\overline{Sp},\mathcal{H)\simeq (}\bigotimes \overline{%
\rho }_{\alpha },\prod \overline{Sp}_{\alpha },\bigotimes \mathcal{H}%
_{\alpha }),
\end{equation*}%
where $(\overline{\rho }_{\alpha },\overline{Sp}_{\alpha },\mathcal{H}%
_{\alpha })$ is the Weil representation associated with the Heisenberg
representation\ $\overline{\pi }_{\alpha }$.
\end{theorem}

\begin{remark}
As before, we can state an equivalent result using the Jacobi groups $%
J=Sp\ltimes H$ and $\overline{J}=\overline{Sp}\ltimes \overline{H}$. We have
a decomposition $\overline{J}=\prod \overline{J}_{\alpha },$ where $%
\overline{J}_{\alpha }=\overline{Sp}_{\alpha }\ltimes \overline{H}_{\alpha }$%
. Let $\tau $ be the Heisenberg--Weil representation of $J$ associated with a
character $\psi $ of the center $Z(J)$ $($note that $Z(J)=Z(H))$. Then the
pullback $\overline{\tau }=$ $\iota ^{\ast }\tau $ is isomorphic to $%
\bigotimes \overline{\tau }_{\alpha }$, where $\overline{\tau }_{\alpha }$
is the Heisenberg--Weil representation of $\overline{J}_{\alpha }$,
associated with the character $\overline{\psi }_{\alpha }=\psi \circ
\textrm{Tr}_{K_{\alpha }/k}$ of the center $Z(\overline{J}_{\alpha })$.
\end{remark}

\subsection{Application to multiplicities\label{HDM}}
Let us specialize to the case where the filed $k$ is a finite field of odd
characteristic. Let $T$ $\subset Sp$ be a maximal torus. The torus $T$ acts,
via the Weil representation $\rho $, on the space $\mathcal{H}$, decomposing
it into a direct sum of $T$-character spaces $\mathcal{H=}%
\bigoplus \limits_{\chi :T\rightarrow
\mathbb{C}
^{\times }}\mathcal{H}_{\chi }.$ Consider the problem of determining the
multiplicities $m_{\chi }=\dim (\mathcal{H}_{\chi })$. Using Lemma \ref%
{technical}, we have (see (\ref{IOG})) a canonical decomposition of $T\ $%
\begin{equation}
T=\prod T_{\alpha },  \label{dec}
\end{equation}%
where each of the tori $T_{\alpha }$ coincides with a maximal torus inside $%
\overline{Sp}\simeq SL_{2}(K_{\alpha }),$ for some field extension $%
K_{\alpha }\supset k.$ In particular, by (\ref{dec}) we have a decomposition
\begin{equation}
\mathcal{H}_{\chi }\mathcal{=}\bigotimes \limits_{\chi _{\alpha }:T_{\alpha
}\rightarrow
\mathbb{C}
^{\times }}\mathcal{H}_{\chi _{\alpha }},  \label{decchar}
\end{equation}%
where $\chi =\prod \chi _{\alpha }:\prod T_{\alpha }\rightarrow
\mathbb{C}
^{\times }$. Hence, by Theorem \ref{Weildecomp}, and the result about the
multiplicities in the two-dimensional case (see Theorem (\ref%
{Multiplicities})), we can compute the integer $m_{\chi }$ as follows$.$
Denote by $\sigma _{\alpha \text{ }}$the quadratic character of $T_{\alpha }$
(note that by Theorem (\ref{Multiplicities}) the quadratic character $%
\sigma _{\alpha }$ cannot appear in the decomposition (\ref{decchar}) if
the torus $T_{\alpha }$ is inert).

\begin{theorem}
\label{MF} We have
\begin{equation*}
m_{\chi }=2^{l},
\end{equation*}%
where $l=\# \{ \alpha :$ $\chi _{\alpha }=$ $\sigma _{\alpha }\}.$
\end{theorem}

\section{Bounds on Higher-Dimensional Exponential Sums\label{B}}
In this section we present an application of the self-reducibility
technique to bound higher-dimensional exponential sums attached to tori in $Sp=Sp(V,\omega ),$ where $(V,\omega )$
is a $2N$-dimensional symplectic vector space over the finite field $\mathbb{%
F}_{p},$ $p\neq 2$. These exponential sums originated from the theory of
quantum chaos (see Sections \ref{HBM} and \ref{HQUE}). Let $(\tau ,J,%
\mathcal{H)}$ be the Heisenberg--Weil representation associated with a
central character $\psi :Z(J)=Z(H)\rightarrow
\mathbb{C}
^{\times }$. Recall that $J=Sp\ltimes H$, and $\tau $ is obtained as a
semi-direct product $\tau =\rho \ltimes \pi $ of the Weil representation $%
\rho $ and the Heisenberg representation $\pi $. Consider a maximal torus $%
T\subset Sp.$ The torus $T$ acts semisimply on $\mathcal{H}$, decomposing it
into a direct sum of character spaces $\mathcal{H=}\bigoplus \limits_{\chi
:T\rightarrow
\mathbb{C}
^{\times }}\mathcal{H}_{\chi }.$ We shall study common eigenstates $\Psi \in
$ $\mathcal{H}_{\chi }$. In particular, we will be interested in estimating
matrix coefficients of the form $\left \langle \Psi |\pi (\xi )\Psi
\right \rangle $ where $\xi \in V$ is not contained in any proper $T$%
-invariant subspace.  It will be convenient to assume first that the torus $T$ is
completely inert (i.e., acts irreducibly on $V$). In this case one can show (see Theorem \ref%
{MF}) that $\dim \mathcal{H}_{\chi }=1$ for every $\chi$. Below we sketch a proof of the following
estimate.
\begin{theorem}
\label{KR1} For $\xi \in V$ which is not contained in any proper $T$%
-invariant subspace, we have
\begin{equation*}
\left \vert \left \langle \Psi |\pi (\xi )\Psi \right \rangle \right \vert \leq
\frac{2+o(1)}{\sqrt{p}^{N}}.
\end{equation*}
\end{theorem}

Let us explain way it is not easy to get such a bound by a direct calculation.
Choosing a Schr\"{o}dinger realization (see Remark \ref{Schrodinger}), we can identify $\mathcal{H}=\mathbb{C}(\mathbb{F}_{p}^{N})$.
Under this identification, the matrix coefficient is equivalent
to a sum
\begin{equation}
\left \langle \Psi |\pi (\xi )\Psi \right \rangle =\sum\limits_{x\in \mathbb{F%
}_{p}^{N}}\Psi (x)e^{\frac{2\pi i}{p}\xi _{+}x}\overline{\Psi }(x+\xi _{-}).
\label{integral1}
\end{equation}%
In this respect two problems are encountered. First, it is not easy to
describe the eigenstates $\Psi $. Second, the sum (\ref{integral1}) is
a high-dimensional exponential sum, which is known to be hard to analyze
using standard techniques.

Interestingly enough, representation theory suggests a remedy for both
problems. Our strategy will be to interpret the matrix coefficient $%
\left \langle \Psi |\pi (\xi )\Psi \right \rangle $ in representation
theoretic terms, and then to show, using the self-reducibility technique,
that (\ref{integral1}) is equivalent to a $1$-dimensional sum over $%
\mathbb{F}_{q},$ $q=p^{N}.$

\subsubsection{Representation theory and dimensional reduction of (\ref{integral1})}
The torus $T$ acts irreducibly on the vector space $V$. Invoking the result of Section
\ref{particular}, there exists a canonical symplectic module structure $(K,V,\overline{\omega })$
associated to $T$. Recall that in this particular case
the algebra $K$ is in fact a field, and $\dim _{K}V=2$ (in our case $K=%
\mathbb{F}_{q},$ where $q=p^{N}$ ). Let $\overline{J}=\overline{Sp}\ltimes
\overline{H}$ be the Jacobi group associated to the (two-dimensional)
symplectic vector space $(V,\overline{\omega })$ over $K$. There exists a
natural homomorphism $\iota :\overline{J}\rightarrow J$. Invoking the
results of\ Section \ref{particular1}, the pullback $\overline{\tau }=$ $%
\iota ^{\ast }\tau $ is the Heisenberg--Weil representation of $\overline{J}$%
, i.e., $\overline{\tau }=\overline{\rho }\ltimes \overline{\pi }$.

Let $\Psi \in \mathcal{H}_{\chi }$. Denote by $P_{\chi }$ the orthogonal
projector on the vector space $\mathcal{H}_{\chi }$. We can write $P_\chi$
in terms of the Weil representation $\overline{\rho }$
\begin{equation}
P_{\chi }=\frac{1}{|T|}\sum \limits_{B\in T}\chi ^{-1}(B)\overline{\rho }(B).
\label{project1}
\end{equation}%
Since $\dim \mathcal{H}_{\chi }=1$ (Theorem \ref{MF}) we realize that
\begin{equation}
\left \langle \Psi |\pi (\xi )\Psi \right \rangle =\textrm{Tr}(P_{\chi }\pi (\xi )).
\label{MCET}
\end{equation}%
Substituting (\ref{project1}) in (\ref{MCET}), we can write
\begin{equation*}
\left \langle \Psi |\pi (\xi )\Psi \right \rangle =\frac{1}{|T|}%
\sum \limits_{B\in T}\chi ^{-1}(B)\textrm{Tr}(\overline{\rho }(B)\pi (\xi )).
\end{equation*}

Noting that $\textrm{Tr}(\overline{\rho }(B)\pi (\xi ))$ is nothing other than the character
$\textrm{ch}_{\overline{\tau }}(B\cdot \xi )$ of the Heisenberg--Weil representation $%
\overline{\tau }.$ and that $|T|=p^N + 1$, we deduce that it is enough to prove that

\begin{equation}
\left \vert \sum \limits_{B\in T}\chi ^{-1}(B)\textrm{ch}_{\overline{\tau }}(B\cdot
\xi )\right \vert \leq 2\sqrt{q},  \label{onedim1}
\end{equation}%
where $q=p^{N}$. Now, note that the left-hand side of (\ref{onedim1}) is a one-dimensional exponential sum over $\mathbb{F}_{q}$, which is defined completely in terms of the two-dimensional Heisenberg--Weil
representation $\overline{\tau }$. Estimate (\ref{onedim1}) is then a
particular case of the following theorem, proved in \cite{GH3}.

\begin{theorem}
\label{twodim1}Let $(V,\omega )$ be a two-dimensional symplectic vector
space over a finite field $k=\mathbb{F}_{q}$, and $(\tau ,J,\mathcal{H)}$ be
the corresponding Heisenberg--Weil representation. Let $T\subset Sp$ be a
maximal torus. We have the following estimate
\begin{equation}
\left \vert \sum \limits_{B\in T}\chi (B)\textrm{ch}_{\tau }(B\cdot \xi )\right \vert
\leq 2\sqrt{q},  \label{2dim_estimate1}
\end{equation}%
where $\chi $ is a character of $T$, and $0\neq \xi \in V$ is not an
eigenvector of $T$.
\end{theorem}

\subsection{General case}
In this subsection we state and prove the analogue of Theorem \ref{KR1},
where we drop the assumption of $T$ being completely inert. In what follows,
we use the results of Subsections \ref{general} and \ref{general1}.

Let $(K,V,\overline{\omega })$ be the symplectic module structure associated
with the torus $T$. The algebra $K$ is no longer a field, but decomposes
into a direct sum of fields $K=\bigoplus \limits_{\alpha \in \Xi }K_{\alpha }
$. We have canonical decompositions
\begin{eqnarray*}
(V,\omega ) =\bigoplus (V_{\alpha },\omega _{\alpha }), \ \ \ \
(V,\overline{\omega }) =\bigoplus (V_{\alpha },\overline{\omega }_{\alpha
}).
\end{eqnarray*}

Recall that $V_{\alpha }$ is a two-dimensional vector space over the field $%
K_{\alpha }$. The Jacobi group $\overline{J}$ decomposes into $\overline{J}%
=\prod \overline{J}_{\alpha },$ where $\overline{J}_{\alpha }=\overline{Sp}%
_{\alpha }\ltimes \overline{H}_{\alpha }$ is the Jacobi group associated to $%
(V_{\alpha },\overline{\omega }_{\alpha }).$ The pullback $(\overline{\tau }%
=$ $\iota ^{\ast }\tau ,\overline{J},\mathcal{H)}$ decomposes into a tensor
product $(\bigotimes \overline{\tau }_{\alpha },\prod \overline{J}_{\alpha
},\bigotimes \mathcal{H}_{\alpha }),$ where $\overline{\tau }_{\alpha }$ is
the Heisenberg--Weil representation of $\overline{J}_{\alpha }$. The torus
$T$ decomposes into $T=\prod T_{\alpha },$ where \bigskip $T_{\alpha }$ is
a maximal torus in $\overline{Sp}_{\alpha }.$ Consequently, the character $\chi :T\rightarrow
\mathbb{C}
^{\times }$ decomposes into a product $\chi =\Pi \chi _{\alpha }:\Pi
T_{\alpha }\rightarrow
\mathbb{C}
^{\times }$, and the space $\mathcal{H}_{\chi }$ decomposes into a tensor
product
\begin{equation}
\mathcal{H}_{\chi }=\bigotimes\mathcal{H}_{\chi _{\alpha }}.  \label{ESD}
\end{equation}

It follows from the above decomposition that it is enough to estimate matrix
coefficients with respect to \textit{pure tensor} eigenstates, i.e., eigenstates $%
\Psi $ of the form $\Psi =\bigotimes \Psi _{\alpha }$, where $\Psi _{\alpha
}\in \mathcal{H}_{\chi _{\alpha }}$. For a vector of the form $\xi
=\bigoplus \xi _{\alpha }$, we have
\begin{equation}
\left \langle \bigotimes \Psi _{\alpha }|\pi (\xi )\bigotimes \Psi _{\alpha
}\right \rangle =\prod \left \langle \Psi _{\alpha }|\pi (\xi _{\alpha })\Psi
_{\alpha }\right \rangle .  \label{MultipF}
\end{equation}

Hence, we need to estimate the matrix coefficients $\left \langle \Psi
_{\alpha }|\pi (\xi _{\alpha })\Psi _{\alpha }\right \rangle $, but these are
defined in terms of the two-dimensional Heisenberg--Weil representation $%
\overline{\tau }_{\alpha }$. In addition, we recall the assumption that the
vector $\xi \in V$ is not contained in any proper $T$-invariant subspace.
This condition in turn implies that no summand $\xi _{\alpha }$ is an
eigenvector of $T_{\alpha }.$ Hence, we can use Lemma \ref{twodim1},
obtaining
\begin{equation}
\left \vert \left \langle \Psi _{\alpha }|\pi (\xi _{\alpha })\Psi _{\alpha
}\right \rangle \right \vert \leq 2/\sqrt{p}^{[K_{\alpha }:\mathbb{F}_{p}]}.
\label{Bound}
\end{equation}
Consequently, using $\left( \text{\ref{MultipF}}\right) $ and (\ref{Bound}%
) we obtain
\begin{equation*}
\left \vert \left \langle \bigotimes \Psi _{\alpha }|\pi (\xi )\bigotimes
\Psi _{\alpha }\right \rangle \right \vert \leq 2^{\left \vert \Xi \right \vert
}/\sqrt{p}^{\sum [K_{\alpha }:\mathbb{F}_{p}]}=2^{\left \vert \Xi
\right \vert }/\sqrt{p}^{[K:\mathbb{F}_{p}]}=2^{\left \vert \Xi \right \vert }/%
\sqrt{p}^{N}.
\end{equation*}

Recall that the number $r_{p}=\left \vert \Xi \right \vert $ is called the
symplectic rank of the torus $T$. The main application of the
self-reducibility property, presented in these notes, is summarized in the
following theorem.

\begin{theorem}
\label{KR_general1}Let $(V,\omega )$ be a $2N$-dimensional vector space over
the finite field $\mathbb{F}_{p}$, and $(\tau ,J,\mathcal{H)}$ the
corresponding Heisenberg--Weil representation. Let $\Psi \in \mathcal{H}%
_{\chi }$ be a unit $\chi $-eigenstate with respect to a maximal torus $%
T\subset Sp$. We have the following estimate:
\begin{equation*}
\left \vert \left \langle \Psi |\pi (\xi )\Psi \right \rangle \right \vert \leq
\frac{m_{\chi }\cdot (2+o(1))^{r_{p}}}{\sqrt{p}^{N}},
\end{equation*}%
where $1\leq r_{p}\leq N$ is the symplectic rank of $T$, $m_{\chi }=\dim
\mathcal{H}_{\chi }$, and $\xi \in V$ is not contained in any $T$-invariant
subspace.
\end{theorem}

\section{The Hannay--Berry model\label{HBM}}
We shall proceed to describe the higher-dimensional Hannay--Berry model of
quantum mechanics on toral phase spaces. This model plays an important role
in the mathematical theory of quantum chaos as it serves as a model where
general phenomena, which are otherwise treated only on a heuristic basis,
can be rigorously proven.

\subsection{The classical phase space}
Our classical phase space is the $2N$-dimensional symplectic torus $(\mathbb{%
T},\omega ).$ We denote by $\Gamma $ the group of linear symplectic
automorphisms of $\mathbb{T}$. Note that $\Gamma \simeq Sp(2N,%
\mathbb{Z}
).$ On the torus $\mathbb{T}$ we consider an algebra of complex functions
(observables) $\mathcal{A}=\mathcal{F}(\mathbb{T)}$. We denote by $\Lambda\simeq\mathbb{Z}^{2N}$
the lattice of characters (exponents) of $\mathbb{T}$. The form $\omega $
induces a skew-symmetric form on $\Lambda $, which we denote also by $\omega
$, and we assume it takes integral values\ on $\Lambda $ and is normalized so
that $\int_{\mathbb{T}}|\omega |^{N}=1.$

\subsection{The classical mechanical system}
We take our classical mechanical system to be of a very simple nature. Let $%
A\in \Gamma $ be a generic element (see Definition \ref{qergodic}), i.e., $A$
is regular and admits no invariant co-isotropic sub-tori. The last condition
can be equivalently restated in dual terms, namely, requiring that $A$
admits no invariant isotropic subvectorspaces in $\Lambda _{%
\mathbb{Q}
}=\Lambda \otimes _{%
\mathbb{Z}
}%
\mathbb{Q}
$. The element $A$ generates, via its action as an automorphism $A:\mathbb{%
T\longrightarrow T},$\ a discrete time dynamical system.

\subsection{Quantization}
Before we employ the formal model, it is worthwhile to discuss the general
phenomenological principles of quantization which are common to all models.
Principally, quantization is a protocol by which one associates a Hilbert
space $\mathcal{H}$ to the classical phase space, which in our case is the
torus $\mathbb{T};\ $In addition, the protocol gives a rule
\begin{equation*}
f\rightsquigarrow Op(f):\mathcal{H\rightarrow H},
\end{equation*}%
by which one associates an operator on the Hilbert space to every classical
observable, i.e., a function $f\in \mathcal{F}(\mathbb{T}).$ This rule
should send a real function into a self-adjoint operator. In addition, in
the presence of classical symmetries which in our case are given by the
group $\Gamma $, the Hilbert space $\mathcal{H}$ should support a
(projective unitary) representation $\Gamma \rightarrow PGL(\mathcal{H})$,
$$
\gamma \mapsto U(\gamma):\mathcal{H\rightarrow H},
$$
which is compatible with the quantization rule $Op(\cdot )$.

More precisely, quantization is not a single protocol, but a one-parameter
family of protocols, parameterized by a parameter $\hslash $ called the
Planck constant. Accepting these general principles, one searches for a
formal model by which to quantize. In this work we employ a model called the
\textit{non-commutative torus} model.

\subsection{The non-commutative torus model\label{RM}}
Denote by $\mathcal{A}$ the algebra of trigonometric polynomials on $\mathbb{%
T}$, i.e., $\mathcal{A}$ consists of functions $f$ which are a finite
linear combinations of characters. We shall construct a one-parametric
deformation of $\mathcal{A}$ called the non-commutative torus \cite{Co, Ri}.

Let $\hbar =1/p,$ where $p$ is an odd prime number, and consider the
additive character $\psi :\mathbb{F}_{p}\longrightarrow {\mathbb{C}}^{\times
},\; \psi (t)=e^{\frac{2\pi it}{p}}$. We give here the following presentation
of the algebra $\mathcal{A}_{\hbar }$. Let $\mathcal{A}_{\hslash }$ be the
free non-commutative $%
\mathbb{C}
$-algebra generated by the symbols $s(\xi )$, $\xi \in \Lambda ,$ and the relations
\begin{equation}
s(\xi )s(\eta )=\psi (\tfrac{1}{2}\omega (\xi ,\eta ))s(\xi +\eta ).
\label{CCR}
\end{equation}%
Here we consider $\omega $ as a map $\omega :\Lambda \times \Lambda
\longrightarrow \mathbb{F}_{p}$.

Note that $\mathcal{A}_{\hslash }$ satisfies the following properties:

\begin{itemize}
\item As a vector space $\mathcal{A}_{\hslash }$ is equipped with a natural
basis $\ s(\xi )$, $\xi \in \Lambda $. Hence we can identify the vector
space $\mathcal{A}_{\hslash }$ with the vector space $\mathcal{A}$ for each
value of $\hslash$,
\begin{equation}
\mathcal{A}_{\hslash }\simeq \mathcal{A}.  \label{ident}
\end{equation}

\item Substituting $\hslash =0$ we have $\mathcal{A}_{0}=\mathcal{A}$.
Hence, we see that indeed $\mathcal{A}_{\hslash }$ is a deformation of the
algebra of (algebraic) functions on $\mathbb{T}$.

\item The group $\Gamma $ acts by automorphisms on the algebra $\mathcal{A}%
_{\hslash }$, via $\gamma\cdot s(\xi)=s(\gamma \xi)$, where $\gamma \in \Gamma $
and $\xi\in \Lambda$. This action induces an action of $\Gamma $
on the category of representations of $\mathcal{A}_{\hslash }$, taking a
representation $\pi $ and sending it to the representation $\pi ^{\gamma }$,
where $\pi ^{\gamma }(f)=\pi (\gamma f)$, $f\in \mathcal{A}_{\hslash }$.
\end{itemize}

Using the identification $\left( \text{\ref{ident}}\right) ,$ we can
describe a choice for the quantization of the functions. We just need to
pick a representation of the quantum algebra $\mathcal{A}_{\hslash }.$ But
what representation to pick? It turns out that, we have a canonical choice.
All the irreducible algebraic representations of $\mathcal{A}_{\hslash }$
are classified \cite{GH2} and each of them is of dimension $p^N.$ We have

\begin{theorem}[Invariant representation \protect \cite{GH2}]
\label{fixrep}Let $\hslash =1/p$ where $p$ is a prime number. There exists a
unique (up to isomorphism) irreducible representation $\pi :\mathcal{A}_{\hslash }\rightarrow
\textrm{End}(\mathcal{H}_{\hslash })$ which is fixed by the action of $\Gamma $.
Namely, $\pi ^{\gamma }$ is isomorphic to $\pi $ for every $\gamma \in
\Gamma $.
\end{theorem}

Let $(\pi ,\mathcal{A}_{\hslash },\mathcal{H)}$ be a representative of the special
representation defined in Theorem \ref{fixrep}. For every element $\gamma \in
\Gamma $ we have an isomorphism $\widetilde{\rho }(\gamma ):\mathcal{
H\rightarrow H}$ intertwining the representations $\pi $ and $\pi ^{\gamma }$
, namely, it satisfies $\widetilde{\rho }(\gamma )\pi (f)\widetilde{\rho }(\gamma )^{-1}=\pi (\gamma f),$
for every $f\in \mathcal{A}_{\hslash }$ and $\gamma \in \Gamma $. The isomorphism $\widetilde{\rho }(\gamma )$ is not
unique but unique up to a scalar (this is a consequence of Schur's lemma
and the fact that $\pi $ and $\pi ^{\gamma }$ are irreducible
representations). It is easy to realize that the collection $\{ \widetilde{%
\rho }(\gamma )\}$ constitutes a projective representation $\widetilde{\rho }%
:\Gamma \rightarrow PGL(\mathcal{H)}$. Let $\hslash =1/p$ where $p$ is an
odd prime $\neq 3.$ We have the following linearization theorem (cf. \cite%
{GH1, GH3})

\begin{theorem}[Linearization]
\label{factorization}The projective representation $\widetilde{\rho }$ can
be linearized uniquely to an honest representation $\rho :\Gamma
\rightarrow GL(\mathcal{H})$ that factors through the finite quotient group $%
Sp\simeq Sp(2N,\mathbb{F}_{p})$.
\end{theorem}

\begin{remark}
The representation $\rho :Sp\rightarrow GL(\mathcal{H)}$ is the celebrated
Weil representation, here obtained via quantization of the torus.
\end{remark}

\subsection{The quantum dynamical system}
Recall that we started with a classical dynamic on $\mathbb{T}$, generated
by a generic (i.e., regular with no non-trivial invariant co-isotropic sub-tori) element $%
A\in \Gamma $. Using the Weil representation, we can associate to $A$ the unitary operator $\rho (A):\mathcal{H\rightarrow H}$, which constitutes the generator of discrete time quantum dynamics.
We would like to study the $\rho(A)$-eigenstates
$$
\rho (A)\Psi =\lambda \Psi,
$$
which satisfy additional symmetries. This we do in the next section.

\section{The Hecke quantum unique ergodicity theorem\label{HQUE}}
It turns out that the operator $\rho (A)$ has degeneracies namely, its eigenspaces might be
extremely large. This is manifested in the existence of a group of hidden
symmetries commuting with $\rho (A)$ (note that classically the group of linear symplectomorphisms
of $\mathbb{T}$ that commute with $A$, i.e., $\mathbf{T}_{A}(%
\mathbb{Z}
),$ does not contribute much to the harmonic analysis of $\rho (A)$). These symmetries can be computed.
Indeed, let $T_{A}=Z(A,Sp),$ be the centralizer of the element $A$ in the
group $Sp$. Clearly $T_{A}$ contains the cyclic group $\langle A \rangle$ generated by the
element $A,$ but it often happens that $T_{A}$ contains additional elements.
The assumption that $A$ is regular (i.e., has distinct eigenvalues) implies that for sufficiently large
$p$ the group $T_{A}$ consists of the $\mathbb{F}_{p}$-rational points of a
maximal torus $\mathbf{T}_{A} \subset \mathbf{Sp}$, i.e., $T_{A}=\mathbf{T}_{A}(\mathbb{F}_{p})$ (more precisely, $p$ large enough so that it does not divides the discriminant of $A$). The group $T_{A}$ is called the
\textit{Hecke }torus. It acts semisimply on $\mathcal{H}$, decomposing it
into a direct sum of character spaces $\mathcal{H=}\bigoplus \limits_{\chi
:T_{A}\rightarrow
\mathbb{C}
^{\times }}\mathcal{H}_{\chi }.$ We shall study common eigenstates $\Psi \in
$ $\mathcal{H}_{\chi },$ which we call \textit{Hecke eigenstates }and will
be assumed to be normalized so that $\left \Vert \Psi \right \Vert _{%
\mathcal{H}}=1$. In particular, we will be interested in estimating matrix
coefficients of the form $\left \langle \Psi |\pi (f)\Psi \right \rangle ,$
where $f$ $\in \mathcal{A}$ is a classical observable on the torus $\mathbb{T%
}$ (see Subsection \ref{RM}). We will call these matrix coefficients \textit{%
Hecke--Wigner distributions}. It will be convenient for us to start with the
following case.

\subsection{The strongly generic case \label{TSGC}}
Let us assume first that the automorphism $A$ acts on $\mathbb{T}$ with no
invariant sub-tori. In dual terms, this means that the element $A$ acts
irreducibly on the $%
\mathbb{Q}
$-vector space $\Lambda _{%
\mathbb{Q}
}=\Lambda \otimes _{%
\mathbb{Z}
}%
\mathbb{Q}
.$

We denote by $r_{p}$ the symplectic rank of $T_{A}$, i.e., $r_{p}=|\Xi |$
where $\Xi =\Xi (T_{A})$ is the symplectic type of $T_{A}$ (see Definition %
\ref{type}). By definition we have $1\leq r_{p}\leq N$ (for example, we get the two extreme cases:
$d_{p}=1$ when the torus $T_{A}$ acts irreducibly on $V\simeq\mathbb{F}_p^{2N}$, and $d_{p}=N$ when $T_{A}$ splits).
We have
\begin{theorem}
\label{SGC} Consider a non-trivial exponent $0\neq \xi \in \Lambda $ and a
sufficiently large prime number $p.$ Then for every normalized Hecke
eigenstate $\Psi \in $ $\mathcal{H}_{\chi }$ the following bound holds:
\begin{equation}
\left \vert \left \langle \Psi |\pi (\xi )\Psi \right \rangle \right \vert \leq
\frac{m_{\chi }\cdot 2^{r_{p}}}{\sqrt{p}^{N}},  \label{SGCF}
\end{equation}
where $m_\chi = \dim(\mathcal{H}_\chi)$.
\end{theorem}
The lattice $\Lambda $ constitutes a basis for $\mathcal{A}$, hence, using the
bound (\ref{SGCF}) we obtain

\begin{corollary}[Hecke quantum unique ergodicity---strongly generic case]
Consider an observable $f\in \mathcal{A}$ and a sufficiently large prime
number $p$. For every normalized Hecke eigenstate $\Psi $ we have
\begin{equation*}
\left \vert \left \langle \Psi |\pi (f)\Psi \right \rangle -\int \limits_{%
\mathbb{T}}fd\mu \right \vert \leq \frac{C_{f}}{\sqrt{p}^{N}},
\end{equation*}%
where $\mu =|\omega |^{N}$ is the corresponding volume form and $C_{f}$ is
an explicit computable constant which depends only on the function $f.$
\end{corollary}

\begin{remark}
In Subsection \ref{St} we will elaborate on the distribution of the
symplectic rank $r_{p}$ $\left( \text{\ref{SGCF}}\right) $ and in Subsection %
\ref{GGC} the more general statements where $A\in \Gamma $ is any generic
element $($see Definition\  \ref{qergodic}$)$ will be stated and proved.
\end{remark}

\subsubsection{Proof of Theorem \protect \ref{SGC}}
The proof is by reduction to the bound on the Hecke--Wigner distributions
obtained in Section \ref{B}, namely reduction to Theorem \ref{KR_general1}.
Our first goal is to interpret the Hecke--Wigner distribution $\left \langle
\Psi |\pi (\xi )\Psi \right \rangle $ in terms of the Heisenberg--Weil
representation.

\textbf{Step 1.} \textit{Replacing the non-commutative torus by the finite
Heisenberg group.} Note that the Hilbert space $\mathcal{H}$ is a
representation space of both the algebra $\mathcal{A}_{\hslash }$ and the
group $Sp$. We will show next that the representation $(\pi ,\mathcal{A}%
_{\hslash },\mathcal{H})$ is \textit{equivalent} to the Heisenberg representation
of some finite Heisenberg group. The representation $\pi $ is determined by
its restriction to the lattice $\Lambda $. However, the restriction
\begin{equation*}
\pi _{|\Lambda }:\Lambda \rightarrow GL(\mathcal{H)},
\end{equation*}%
is not multiplicative and in fact constitutes (see Formula (\ref{CCR})) a
projective representation of the lattice given by
\begin{equation}
\pi (\xi )\pi (\eta )=\psi (\tfrac{1}{2}\omega (\xi ,\eta ))\pi (\xi +\eta ).
\label{CRF}
\end{equation}%
It is evident from $\left( \text{\ref{CRF}}\right) $ that the map $\pi
_{|\Lambda }$ factors through the quotient $\mathbb{F}_{p}$-vector space $V$
\begin{equation}
\Lambda \rightarrow V=\Lambda /p\Lambda \rightarrow GL(\mathcal{H)}.
\label{Fac}
\end{equation}
The vector space $V$ is equipped with a symplectic structure $\omega $
obtained via specialization of the corresponding form on $\Lambda .$ Let $%
H=H(V,\omega)$ be the Heisenberg group associated with $(V,\omega )$. Recall that
as a set $H=V\times $ $\mathbb{F}_{p}$ and the multiplication is given by
\begin{equation}
(v,z)\cdot (v^{\prime },z^{\prime })=(v+v^{\prime },z+z^{\prime }+\tfrac{1}{2%
}\omega (v,v^{\prime })).  \label{MR}
\end{equation}%
From formula (\ref{CRF}), the factorization (\ref{Fac}), and the
multiplication rule (\ref{MR}) we learn that the map $\pi :V\rightarrow GL(%
\mathcal{H)},$ given by (\ref{Fac}), lifts to an honest representation of
the Heisenberg group $\pi :H\rightarrow GL(\mathcal{H)}.$
Finally, the pair $(\rho ,\pi )$, where $\rho $ is the Weil representation
obtained using quantization of the torus (see Theorem \ref{factorization})
glues into a single representation $\tau =\rho \ltimes \pi $ of the Jacobi
group $J=Sp\ltimes H$, which is of course nothing other than the Heisenberg--Weil
representation
\begin{equation}
\tau :J\rightarrow GL(\mathcal{H)}.  \label{HWR}
\end{equation}%
Having the Heisenberg--Weil representation at our disposal we proceed to
\smallskip

\textbf{Step 2.}\textit{Reformulation. }Let $\mathbf{V}$ and $\mathbf{T}%
_{A} $ be the algebraic group scheme defined over $%
\mathbb{Z}
$ so that $\Lambda =\mathbf{V(%
\mathbb{Z}
)}$ and for every prime $p$ we have $V=\mathbf{V(\mathbb{F}_{p})}$ and $%
T_{A} $ $=\mathbf{T}_{A}(\mathbb{F}_{p}).$ In this setting for every prime
number $p$ we can consider the lattice element $\xi \in \Lambda $ as a
vector in the $\mathbb{F}_{p}$-vector space $V$.

Let $(\tau ,J,\mathcal{H})$ be the Heisenberg--Weil representation (\ref{HWR}%
) and consider a normalized Hecke eigenstate $\Psi \in \mathcal{H}_{\chi }.$
We need to verify that for a sufficiently large prime number $p$ we have
\begin{equation}
\left \vert \left \langle \Psi |\pi (\xi )\Psi \right \rangle \right \vert \leq
\frac{m_{\chi }\cdot 2^{r_{p}}}{\sqrt{p}^{N}},  \label{ref}
\end{equation}%
where $m_{\chi }$ denotes the multiplicity $m_{\chi }=\dim \mathcal{H}_{\chi
}$ and $r_{p}$ is the symplectic rank of $T_{A}.$ This verification is what
we do next. \smallskip

\textbf{Step 3.} \textit{Verification. }We need to show that we meet the
conditions of Theorem \ref{KR_general1}. What is left to check is that for
sufficiently large prime number $p$ the vector $\xi \in V$ is not contained
in any $T_{A}$-invariant subspace of $V.$ Let us denote by $O_{\xi }$ the
orbit $O_{\xi }=T_{A}\cdot \xi .$ We need to show that for sufficiently
large $p$ we have
\begin{equation}
\textrm{Span}_{\mathbb{F}_{p}}\{O_{\xi }\}=V.  \label{Ir}
\end{equation}

The condition (\ref{Ir}) is satisfied since it holds globally. In more
details, our assumption on $A$ guarantees that it holds for the
corresponding objects over the field of rational numbers $%
\mathbb{Q}
$, i.e., $\textrm{Span}_{%
\mathbb{Q}
}\{ \mathbf{T}_{A}(%
\mathbb{Q}
)\cdot \xi \}=\mathbf{V}\left(
\mathbb{Q}
\right)$. Hence (\ref{Ir}) holds for a sufficiently large prime number $p.$

\subsection{The distribution of the symplectic rank \label{St}}
We would like to compute the asymptotic distribution of the symplectic rank $%
r_{p}$ (\ref{ref}) in the set $\left \{ 1,...,N\right \} $, i.e.,
\begin{equation}
\delta (r)=\lim_{x\rightarrow \infty }\frac{\# \left \{ r_{p}=r\text{ };\text{
}p\leq x\right \} }{\pi (x)},  \label{delta}
\end{equation}
where $\pi (x)$ denotes the number of prime numbers up to $x.$

We fix an algebraic closure $\overline{\mathbb{Q}}$ of the field $\mathbb{Q}$,
and denote by $G$ the Galois group $G=\textrm{Gal}(\overline{%
\mathbb{Q}
}/%
\mathbb{Q}
).$ Consider the vector space $\mathbf{V}=\mathbf{V(}\overline{%
\mathbb{Q}
})$. By extension of scalars the symplectic form $\omega $ on $\mathbf{V}%
\left(
\mathbb{Q}
\right) $ induces a $\overline{%
\mathbb{Q}
}$-linear symplectic form on $\mathbf{V,}$ which we will also denote by $%
\omega .$ Let $\mathbf{T}$ denote the algebraic torus $\mathbf{T=}$ $\mathbf{%
T}_{A}(\overline{%
\mathbb{Q}
}).$ The action of $\mathbf{T}$ on $\mathbf{V}$ is completely reducible,
decomposing it into one-dimensional character spaces $\mathbf{V=}%
\bigoplus \limits_{\chi \in \mathfrak{X}}\mathbf{V}_{\chi }.$

Let $\Theta $ be the restriction of the symplectic transpose $(\cdot
)^{t}:\textrm{End}(\mathbf{V})\rightarrow \textrm{End}(\mathbf{V})$ to $\mathbf{T.}$ The
involution $\Theta $ acts on the set of characters $\mathfrak{X}$ by $\chi
\mapsto \Theta (\chi )=\chi ^{-1}$ and this action is compatible with the
action of the Galois group $G$ on $\mathfrak{X}$ by conjugation $\chi
\mapsto g\chi g^{-1}$, where $\chi \in \mathfrak{X}$ and $g\in G$. This
means (recall that $A$ is strongly generic) that we have a transitive action
of $G$ on the set $\mathfrak{X}/\Theta .$ Consider the kernel $\mathrm{K}%
=\ker (G\rightarrow \textrm{Aut}(\mathfrak{X}/\Theta )),$ and the corresponding
finite Galois group $Q=G/\mathrm{K}.$ Considering $Q$ as a subgroup of $\
\textrm{Aut}(\mathfrak{X}/\Theta )$ we define the cycle number $c(C)$ of a conjugacy
class $C\subset Q$ to be the number of irreducible cycles that compose a
representative of $C.$ By a direct application of the Chebotarev theorem
\cite{C} we get

\begin{proposition}[Chebotarev's theorem]
\label{CT}The distribution $\delta $ $\left( \text{\ref{delta}}\right) $
obeys%
\begin{equation*}
\delta (r)=\frac{\left \vert C_{r}\right \vert }{\left \vert Q\right \vert },
\end{equation*}%
where $C_{r}=\underset{c(C)=r}{\underset{C\subset Q}{\cup }}C.$
\end{proposition}

\subsection{The general generic case \label{GGC}}
Let us now treat the more general case where the automorphism $A$ acts on $%
\mathbb{T}$ in a generic way (Definition \ref{qergodic}). In dual terms,
this means that the torus $\mathbf{T}(%
\mathbb{Q}
)=\mathbf{T}_{A}(%
\mathbb{Q}
)$ acts on the symplectic vector space $\mathbf{V}(%
\mathbb{Q}
)$ $=\Lambda \otimes _{%
\mathbb{Z}
}%
\mathbb{Q}
$ decomposing it into an orthogonal symplectic direct sum%
\begin{equation}
(\mathbf{V}(%
\mathbb{Q}
),\omega )=\bigoplus \limits_{\alpha \in \Xi }(\mathbf{V}_{\alpha }(%
\mathbb{Q}
),\omega _{\alpha }),  \label{Qdecomp}
\end{equation}%
with an irreducible action of $\mathbf{T}(%
\mathbb{Q}
)$ on each of the spaces $\mathbf{V}_{\alpha }(%
\mathbb{Q}
)$. For an exponent $\xi \in \Lambda $ define its support with respect to
the decomposition (\ref{Qdecomp}) by $S_{\xi }=\textrm{Supp}(\xi )=\{ \alpha ;$ $%
P_{\alpha }\xi \neq 0\},$ where $P_{\alpha }:\mathbf{V}(%
\mathbb{Q}
)\rightarrow $ $\mathbf{V}(%
\mathbb{Q}
)$ is the projector onto the space $\mathbf{V}_{\alpha }(%
\mathbb{Q}
)$ and denote by $d_{\xi }$ the dimension $d_{\xi }=\sum\limits_{\alpha \in S_{\xi
}}\dim \mathbf{V}_{\alpha }(%
\mathbb{Q}
).$

\bigskip
The decomposition (\ref{Qdecomp}) induces a decomposition of the
torus $\mathbf{T}(%
\mathbb{Q}
)$ into a product of completely inert tori
\begin{equation}
\mathbf{T}(%
\mathbb{Q}
)=\prod \limits_{\alpha \in \Xi }\mathbf{T}_{\alpha }(%
\mathbb{Q}
).  \label{TDQ}
\end{equation}

Consider now a sufficiently large prime number $p$ and specialize all the
objects involved to the finite filed $\mathbb{F}_{p}$. The Hecke torus $T=$ $%
\mathbf{T}(\mathbb{F}_{p})$ acts on the quantum Hilbert space $\mathcal{H}%
\mathbb{\ }$decomposing it into an orthogonal direct sum $\mathcal{H=}%
\bigoplus \limits_{\chi :T\rightarrow
\mathbb{C}
^{\times }}\mathcal{H}_{\chi }.$ The decomposition (\ref{TDQ}) induces
decompositions on the level of groups of points $T=\prod \limits_{\alpha \in
\Xi }T_{\alpha },$ where \bigskip $T_{\alpha }=$ $\mathbf{T}_{\alpha }(%
\mathbb{F}_{p}),$ on the level of characters $\chi =\underset{\alpha }{\Pi }%
\chi :\prod \limits_{\alpha }T_{\alpha }\rightarrow
\mathbb{C}
^{\times },$ and on the level of character spaces $\mathcal{H}_{\chi
}=\bigotimes \limits_{\alpha }\mathcal{H}_{\chi _{\alpha }}.$

For each torus $T_{\alpha }$ we denote by $r_{p,\alpha }=r_{p}(T_{\alpha })$
its symplectic rank (see Definition \ref{type}) and we consider the integer $%
\left \vert S_{\xi }\right \vert \leq r_{p,\xi }\leq d_{\xi }$ given by $%
r_{p,\xi }=\underset{\alpha \in S_{\xi }}{\Pi }r_{p,\alpha }.$

Let us denote by $m_{\chi _{\xi }}$ the dimension $m_{\chi _{\xi
}}=\sum_{\alpha \in S_{\xi }}\dim \mathcal{H}_{\chi _{\alpha }}.$ Finally,
we can state the theorem for the generic case. We have

\begin{theorem}[Hecke quantum unique ergodicity---generic case]
\label{GCT} Consider a non-trivial exponent $0\neq \xi \in \Lambda $ and a
sufficiently large prime number $p.$ Then for every normalized Hecke
eigenstate $\Psi \in $ $\mathcal{H}_{\chi }$ the following bound holds:
\begin{equation}
\left \vert \left \langle \Psi |\pi (\xi )\Psi \right \rangle \right \vert \leq
\frac{m_{\chi _{\xi }}\cdot 2^{r_{p,\xi }}}{\sqrt{p}^{d_{\xi }}}.
\label{GCF}
\end{equation}
\end{theorem}

Considering the decomposition (\ref{Qdecomp}) we denote by $d$ the dimension
$d=\min_{\alpha }\mathbf{V}_{\alpha }(%
\mathbb{Q}
).$ Since the lattice $\Lambda $ constitutes a basis for the algebra $%
\mathcal{A}$ of observables on $\mathbb{T}$, then using the bound (\ref{GCF})
we obtain

\begin{corollary}
Consider an observable $f\in \mathcal{A}$ and a sufficiently large prime
number $p.$ Then for every normalized Hecke eigenstate $\Psi $ we have
\begin{equation*}
\left \vert \left \langle \Psi |\pi (f)\Psi \right \rangle -\int \limits_{%
\mathbb{T}}fd\mu \right \vert \leq \frac{C_{f}}{\sqrt{p}^{d}},
\end{equation*}%
where $\mu =|\omega |^{N}$ is the corresponding volume form and $C_{f}$ is
an explicit computable constant which depends only on the function $f.$
\end{corollary}

The proof of Theorem \ref{GCT} is a straightforward application of Theorem %
\ref{SGC}. Indeed, considering the decomposition (\ref%
{Qdecomp}) of the torus $\mathbf{T}(%
\mathbb{Q}
)$ to a product of completely inert tori $\mathbf{T}_{\alpha }(%
\mathbb{Q}
)$, we may apply the theory developed for the strongly generic case in
Subsection (\ref{TSGC}) to each of the tori $\mathbf{T}_{\alpha }(%
\mathbb{Q}
)$ to deduce Theorem \ref{GCT}.

\begin{remark}
As explained in Subsection $\left( \text{\ref{St}}\right) $ the distribution
of the symplectic rank $r_{p,\xi }$ is determined by the Chebotarev theorem
applied to (now a product of) suitable finite Galois
groups $Q_{\alpha \text{ }}$attached to the tori $\mathbf{T}_{\alpha },$ $%
\alpha \in S_{\xi }$ $\left( \text{\ref{TDQ}}\right) .$
\end{remark}

\begin{remark}
The corresponding quantum unique ergodicity theorem for statistical states
of generic automorphism $A$ of $\mathbb{T}$ $($see Theorem \ref{QUESS}$)$
follows directly from Theorem \ref{GCT}.
\end{remark}


%
%

%
%

\bigskip



\end{document}